\documentclass[twocolumn, trackchanges]{aastex62}

\submitjournal{ApJ}
\usepackage{amsmath}
\usepackage{comment}
\usepackage{cleveref}

\crefformat{section}{\S#2#1#3} 
\crefformat{subsection}{\S#2#1#3}
\crefformat{subsubsection}{\S#2#1#3}

\shortauthors{Rinaldi et al.}

\begin{document}

\title{\bf {Not Just a Dot: the complex UV morphology and underlying properties of Little Red Dots}}

\newcommand{\gsim}{{\;\raise0.3ex\hbox{$>$\kern-0.75em\raise-1.1ex\hbox{$\sim$}}\;}}

\correspondingauthor{Pierluigi Rinaldi}
\email{prinaldi@stsci.edu}

\author[0000-0002-5104-8245]{P. Rinaldi}\altaffiliation{These authors contributed equally to this work.}
\affiliation{Steward Observatory, University of Arizona, 933 North Cherry Avenue, Tucson, AZ 85721, USA}
\affiliation{Kapteyn Astronomical Institute, University of Groningen,
P.O. Box 800, 9700AV Groningen,
The Netherlands
}

\author[0000-0001-8470-7094]{N. Bonaventura}\altaffiliation{These authors contributed equally to this work.}
\affiliation{Steward Observatory, University of Arizona, 933 North Cherry Avenue, Tucson, AZ 85721, USA}
\author[0000-0003-2303-6519]{G. H. Rieke}
\affiliation{Steward Observatory, University of Arizona, 933 North Cherry Avenue, Tucson, AZ 85721, USA}

\author[0000-0002-8909-8782]{S. Alberts}
\affiliation{Steward Observatory, University of Arizona, 933 North Cherry Avenue, Tucson, AZ 85721, USA}

\author[0000-0001-8183-1460]{K. I. Caputi}
\affiliation{Kapteyn Astronomical Institute, University of Groningen,
P.O. Box 800, 9700AV Groningen,
The Netherlands
}
\affiliation{Cosmic Dawn Center (DAWN), Copenhagen, Denmark
}

\author[0000-0003-0215-1104]{W. M. Baker}
\affiliation{Kavli Institute for Cosmology, University of Cambridge, Madingley Road, Cambridge, CB3 OHA, UK}
\affiliation{Cavendish Laboratory - Astrophysics Group, University of Cambridge, 19 JJ Thomson Avenue, Cambridge, CB3 OHE, UK}
\affiliation{DARK, Niels Bohr Institute, University of Copenhagen, Jagtvej 128, DK-2200 Copenhagen, Denmark}

\author[0000-0002-4735-8224]{S. Baum}
\affiliation{Department of Physics and Astronomy, University of Manitoba, Winnipeg, MB R3T 2N2, Canada}

\author[0000-0003-0883-2226]{R. Bhatawdekar}
\affiliation{European Space Agency (ESA), European Space Astronomy Centre (ESAC), Camino Bajo del Castillo s/n, 28692 Villanueva de la Cañada, Madrid, Spain}

\author[0000-0002-8651-9879]{A. J. Bunker}
\affiliation{Department of Physics, University of Oxford, Denys Wilkinson Building, Keble Road, Oxford OX13RH, UK}

\author[0000-0002-6719-380X]{S. Carniani}
\affiliation{Scuola Normale Superiore, Piazza dei Cavalieri 7, I-56126 Pisa, Italy}

\author[0000-0002-9551-0534]{E. Curtis-Lake}
\affiliation{Centre for Astrophysics Research, Department of Physics, Astronomy and Mathematics, University of Hertfordshire, Hatfield AL10 9AB, UK}

\author[0000-0003-2388-8172]{F. D'Eugenio}
\affiliation{Kavli Institute for Cosmology, University of Cambridge, Madingley Road, Cambridge, CB3 0HA, UK}
\affiliation{Cavendish Laboratory, University of Cambridge, 19 JJ Thomson Avenue, Cambridge, CB3 0HE, UK}

\author[0000-0003-1344-9475]{E. Egami}
\affiliation{Steward Observatory, University of Arizona, 933 North Cherry Avenue, Tucson, AZ 85721, USA}

\author[0000-0001-7673-2257]{Z. Ji}
\affiliation{Steward Observatory, University of Arizona, 933 North Cherry Avenue, Tucson, AZ 85721, USA}

\author[0000-0002-9280-7594]{B. D.\ Johnson}
\affiliation{Center for Astrophysics $|$ Harvard \& Smithsonian, 60 Garden St., Cambridge MA 02138 USA}

\author[0000-0003-4565-8239]{K. Hainline}
\affiliation{Steward Observatory, University of Arizona, 933 North Cherry Avenue, Tucson, AZ 85721, USA}

\author[0000-0003-4337-6211]{J. M. Helton}
\affiliation{Steward Observatory, University of Arizona, 933 North Cherry Avenue, Tucson, AZ 85721, USA}

\author[0000-0001-6052-4234]{X. Lin}
\affiliation{Department of Astronomy, Tsinghua University, Beijing 100084, China}
\affiliation{Steward Observatory, University of Arizona, 933 North Cherry Avenue, Tucson, AZ 85721, USA}

\author[0000-0002-6221-1829]{J. Lyu}
\affiliation{Steward Observatory, University of Arizona, 933 North Cherry Avenue, Tucson, AZ 85721, USA}

\author[0000-0000-0000-0000]{Z. Ma}
\affiliation{Steward Observatory, University of Arizona, 933 North Cherry Avenue, Tucson, AZ 85721, USA}

\author[0000-0002-4985-3819]{R. Maiolino}
\affiliation{Kavli Institute for Cosmology, University of Cambridge, Madingley Road, Cambridge, CB3 0HA, UK}
\affiliation{Cavendish Laboratory - Astrophysics Group, University of Cambridge, 19 JJ Thomson Avenue, Cambridge, CB3 0HE, UK}
\affiliation{Department of Physics and Astronomy, University College London, Gower Street, London WC1E 6BT, UK}

\author[0000-0003-4528-5639]{P. G. P{\'e}rez-Gonz{\'a}lez}
\affiliation{Centro de Astrobiolog\'ia (CAB), CSIC–INTA, Cra. de Ajalvir Km.~4, 28850- Torrej\'on de Ardoz, Madrid, Spain}

\author[0000-0002-7893-6170]{M. Rieke}
\affiliation{Steward Observatory, University of Arizona, 933 North Cherry Avenue, Tucson, AZ 85721, USA}

\author[0000-0002-4271-0364]{B. E. Robertson}
\affiliation{Department of Astronomy and Astrophysics, University of California, Santa Cruz, 1156 High Street, Santa Cruz, CA 95064, USA}

\author[0000-0003-4702-7561]{I. Shivaei}
\affiliation{Centro de Astrobiolog\'ia (CAB), CSIC–INTA, Cra. de Ajalvir Km.~4, 28850- Torrej\'on de Ardoz, Madrid, Spain}

\author[0000-0002-9720-3255]{M. Stone}
\affiliation{Steward Observatory, University of Arizona, 933 North Cherry Avenue, Tucson, AZ 85721, USA}

\author[0000-0001-6561-9443]{Y. Sun}
\affiliation{Steward Observatory, University of Arizona, 933 North Cherry Avenue, Tucson, AZ 85721, USA}

\author[0000-0002-8224-4505]{S. Tacchella}
\affiliation{Kavli Institute for Cosmology, University of Cambridge, Madingley Road, Cambridge, CB3 0HA, UK}
\affiliation{Cavendish Laboratory, University of Cambridge, 19 JJ Thomson Avenue, Cambridge, CB3 0HE, UK}

\author[0000-0003-4891-0794]{H. \"Ubler}
\affiliation{Max-Planck-Institut f\"ur extraterrestrische Physik (MPE), Gie{\ss}enbachstra{\ss}e 1, 85748 Garching, Germany}

\author[0000-0003-2919-7495]{C. C. Williams}
\affiliation{NSF National Optical-Infrared Astronomy Research Laboratory, 950 North Cherry Avenue, Tucson, AZ 85719, USA}

\author[0000-0001-9262-9997]{C. N.\ A.\ Willmer}
\affiliation{Steward Observatory, University of Arizona, 933 North Cherry Avenue, Tucson, AZ 85721, USA}

\author[0000-0002-4201-7367]{C. Willott}
\affiliation{NRC Herzberg, 5071 West Saanich Rd, Victoria, BC V9E 2E7, Canada}

\author[0000-0002-1574-2045]{J. Zhang}
\affiliation{Steward Observatory, University of Arizona, 933 North Cherry Avenue, Tucson, AZ 85721, USA}

\author[0000-0003-3307-7525]{Y. Zhu}
\affiliation{Steward Observatory, University of Arizona, 933 North Cherry Avenue, Tucson, AZ 85721, USA}

\begin{abstract}

We analyze 99 photometrically selected Little Red Dots (LRDs) at $z \approx 4$–8 in the GOODS fields, leveraging ultra-deep JADES NIRCam short-wavelength (SW) data. We examine the morphology of 30 LRDs, while the remaining 69 appear predominantly compact, with sizes $\lesssim$400 pc and no extended components even in stacked SW images. However, their unresolved nature may partly reflect current depth limitations, which could prevent the detection of faint diffuse components. Among the 30 morphologically analyzed LRDs, 50\% show multiple associated components, while the rest exhibit highly asymmetric structures, despite appearing as single sources. This diversity in rest-frame UV morphologies may point to interactions or strong internal feedback. We find median stellar masses of $\log_{10}(M_{\star}/M_{\odot}) = 9.07_{-0.08}^{+0.11}$ for pure stellar models with $A_{V} \approx 1.16^{+0.11}_{-0.21}$ mag, and $\log_{10}(M_{\star}/M_{\odot}) = 9.67^{+0.17}_{-0.27}$ for models including AGNs with $A_{V} \approx 2.74^{+0.55}_{-0.71}$ mag, in line with recent studies suggesting higher masses and dust content for AGN-fitted LRDs. NIRSpec spectra are available for 15 sources, six of which are also in the morphological sample. Broad H$\alpha$ is detected in 40\% (FWHM $=1200$–$2900$ km/s), and one source shows broad H$\beta$ emission. Emission line ratios indicate a composite nature, consistent with both AGN and stellar processes. Altogether, these results suggest that LRDs are a mixed population, and their rest-frame UV morphology reflects this complexity. Morphological studies of larger samples could provide a new way to understand what drives their properties and evolution.
\end{abstract}


\section{Introduction}
With its unparalleled sensitivity and angular resolution at infrared (IR) wavelengths, JWST (\citealt{gardner_james_2023}) has
opened new frontiers for exploring the early Universe. Not only has it provided the opportunity to study well-known high-redshift galaxies previously discovered with the {\it Hubble} Space Telescope (HST), such as GN-z11 (\citealt{bouwens_z_2010, oesch_remarkably_2016}), in much greater detail (e.g., \citealt{bunker_jades_2023, maiolino_small_2024, tacchella_jades_2023}), but it has also revealed whole populations of high-redshift galaxies (e.g., \citealt{bradley_high-redshift_2023}). 
One of these  groundbreaking discoveries is the identification of very compact and red sources, initially reported by \citet{labbe_population_2023} and subsequently termed “Little Red Dots” (LRDs) by \citet{matthee_little_2024}.

These sources are characterized by (1) their compactness in the F444W band and (2) their red color at observed wavelengths greater than  $\approx 2\;\mu$m—covered by the Near Infrared Camera (NIRCam; \citealt{rieke_performance_2023}) Long Wavelength (LW) channel. They exhibit a distinct spectral energy distribution (SED) marked by clear Lyman and Balmer breaks, and the characteristic “v-shaped” continuum in the $\lambda$–$\text{f}_\lambda$ plane; i.e., their continua are relatively blue in the rest-ultraviolet but become very red toward the rest-optical. It is worth noting that the location of the “v-shape” feature appears consistently around a rest-frame wavelength of $\approx3600$ {\AA}\;in sufficiently deep spectroscopic observations (e.g., \citealt{furtak_high_2024, wang_rubies_2024, juodzbalis_jades_2024, setton_little_2024}).

At the very beginning, these sources were reported to exhibit uncomfortably large stellar masses ($M_{\star} > 10^{10}\;M_{\odot}$) with a large amount of dust ($A_{V} > 1.5$ mag; see \citealt{labbe_population_2023}). 

Over time, various alternative explanations have been proposed to address these puzzling  results. In some cases, the issue could simply arise from an error in the redshift estimation, as demonstrated, e.g., by \citet{kocevski_hidden_2023} for one of the massive galaxies identified by \citet{labbe_population_2023} and \citet{perez-gonzalez_ceers_2023}.

Another potential solution explains the high $M_{\star}$ values by invoking prominent nebular emission, which could mimic the observed red colors at $\lambda > 2\;\mu$m. Recent studies based on JWST observations have indeed shown high-z galaxies with a prominent (H$\beta$ + [OIII]) complex and/or H$\alpha$ emission lines (e.g.,  \citealt{endsley_jwstnircam_2023, rinaldi_midis_2023,  boyett_extreme_2024, caputi_midis_2024, endsley_star-forming_2024, rinaldi_midis_2024}). Therefore, if strong emission lines are present, the stellar masses could decrease by a factor of 10, as recently reported by \citet{desprez_cdm_2024}.

Because of their puzzling nature, the discovery of the LRDs has triggered, in less than two years, a vast amount of literature (e.g., \citealt{furtak_jwst_2023, labbe_population_2023, labbe_uncover_2023, killi_deciphering_2023, kokorev_uncover_2023, ubler_ga-nifs_2023, akins_cosmos-web_2024, barro_extremely_2024, durodola_exploring_2024, greene_uncover_2024, kocevski_rise_2024, kokorev_census_2024, kokorev_silencing_2024, kokubo_challenging_2024, iani_midis_2024, hainline_investigation_2024, matthee_little_2024, perez-gonzalez_what_2024, williams_galaxies_2024}), leading to one of the most intriguing questions in extragalactic astronomy today: \textit{What is the nature of LRDs?} 

As observations continue, both photometric and spectroscopic, different groups have tried to unveil the true nature of these red and compact sources. Some of them exhibit broad ($\approx1000$ km s$^{-1}$) H$\alpha$ emission lines (\citealt{killi_deciphering_2023, kokorev_uncover_2023, kocevski_hidden_2023, greene_uncover_2024, matthee_little_2024}); therefore, it is widely believed that LRDs could potentially host active galactic nuclei (AGNs). However, their SED model fits can be ambiguous, leaving it unclear whether the emission is primarily driven by an AGN or star formation \citep{barro_extremely_2024}. It is also possible that LRDs are a mixed population with both AGN and star formation-dominated members \cite[e.g.,][]{perez-gonzalez_what_2024}. Interestingly, \citet{perez-gonzalez_what_2024} report that only 17\% of their photometrically selected LRDs present broad spectral components.

It has become evident that the Mid-Infrared Instrument (MIRI; \citealt{wright_mid-infrared_2023}) on board JWST could be a game-changer in studying these objects at IR wavelengths, as it could tip the scale in distinguishing between stellar and AGN emission. Noteworthy are several studies in this regard, including those by \citet{williams_galaxies_2024} and \citet{perez-gonzalez_what_2024}, which suggest that these sources could be either dusty starbursts or obscured AGNs. Particularly, \citet{williams_galaxies_2024}, by making use of the Systematic Mid-infrared Instrument Legacy Extragalactic Survey (SMILES; \citealt{alberts_smiles_2024}) data, found that the average SED of LRDs flattens beyond 5 $\mu$m, indicating the expected turnover of a normal stellar SED at approximately 1.6 $\mu$m rest-frame. Building upon these findings, \citet{perez-gonzalez_what_2024} further concluded that the true nature of the LRDs cannot be uniquely described by a single phenomenon, but rather they are likely to be a non-uniform population of objects, with some being extreme starburst galaxies, some dust-obscured AGNs, and some a combination of both.

In this paper, we propose a new approach to advancing our understanding of LRDs. Although these sources are consistently termed “Little Red Dots” due to their selection as red and compact objects in the F444W band from NIRCam, here we focus on their morphology at observed wavelengths shorter than  $2\;\mu$m, leveraging the superior spatial resolution offered by the NIRCam short-wavelength (SW) channel.  Out of a sample of 99 LRDs $z\approx4-8$, we find that 30 (30\%) are sufficiently extended and have enough signal-to-noise (SNR) per pixel in the rest-frame ultra-violet (UV) for morphological analysis. The remaining 70\% appear predominantly compact ($\lesssim 400$~pc), likely due to their low SNR in each individual band, which prevents the characterization of any extended components, even when stacking the SW bands. While earlier studies have investigated this subject (e.g., \citealt{killi_deciphering_2023, baggen_small_2024}), ours is the first to apply a statistical approach to address it.

Given that our LRD sample spans $z\approx 4-8$, the NIRCam SW bands trace the UV emission from the galaxies, which can include, for example,  star-forming clumps and complexes (e.g., \citealt{guo_clumpy_2015}) with strong contributions from massive O-, B-, and A-type stars (see e.g., \citealt{buta_galaxy_2011, rubinur_study_2024}),  and potentially also the UV continuum emission from an unobscured AGN accretion disk or outflow. This contrasts with the more-common morphology studies conducted in the rest-optical, where galaxy-scale asymmetries and strong disturbances are conventionally attributed to merging activity, as this wavelength range primarily traces the emission from relatively evolved stars.   We therefore complement the rest-UV morphological analysis where possible with additional data from broadband SEDs and spectroscopy, to place our sample of LRDs into the context of previous studies.

The paper is organized as follows. Section \ref{section2} describes the datasets and outlines our sample selection. In Section  \ref{section3}, we explain the methodology for analyzing the UV morphological properties of the selected LRDs. Section \ref{section4} presents the SED fitting configuration used to derive stellar properties and discusses these results. In Section \ref{section5}, we examine the spectral properties of the LRD candidates with NIRSpec spectra, followed by an analysis of the morphology of the LRD showing broad Balmer lines in Section \ref{section6}. Finally, Section \ref{section7} provides a summary and discussion of our findings.

Throughout this paper, we consider a cosmology with $H_{0} = 70\; \rm km\;s^{-1}\;Mpc^{-1}$, $\Omega_{M} = 0.3$, and $\Omega_{\Lambda} =0.7$. All magnitudes are total and refer to the AB system \citep{oke_secondary_1983}. A \citet{kroupa_variation_2001} initial mass function (IMF) is assumed (0.1--100 M$_{\odot}$).

\section{Dataset \& Sample Selection}
\label{section2}
\subsection{Dataset}
In this study, we utilized data from both JWST and HST in the GOODS-North and GOODS-South fields (\citealt{giavalisco_great_2004}; hereafter GOODS-N and GOODS-S).

\subsubsection{NIRCam}
We made use of NIRCam data from JADES/NIRCam Data Release 2 (JADES DR2 -- PIDs: 1180, 1210; PIs.: D. Eisenstein, N. Luetzgendorf; \citealt{eisenstein_overview_2023, eisenstein_jades_2023}), which includes observations from the JWST Extragalactic Medium-band Survey (JEMS -- PID: 1963; PIs: C. C. Williams, S. Tacchella, M. Maseda; \citealt{williams_jems_2023}) for GOODS-S and the First Reionization Epoch Spectroscopically Complete Observations (FRESCO -- PID: 1895; PI: P. Oesch; \citealt{oesch_jwst_2023}) for both GOODS-N and GOODS-S. Additionally, we incorporated NIRCam data from JADES Data Release 3 (DR3) for GOODS-N (\citealt{deugenio_jades_2024}).

The JADES/NIRCam data allow us to cover a wide range in wavelengths ($\approx 1\mu\text{m} - 5\mu\text{m}$). Specifically, the dataset in GOODS-N allows us to make use of 11 NIRCam bands (both medium- and broad bands; $0.9\mu\text{m} - 4.44\mu\text{m}$), while the dataset in GOODS-S allows us to make use of 14 bands (both medium- and broad bands; $0.9\mu\text{m} - 4.80\mu\text{m}$).

We estimate a 5$\sigma$ depth ranging from 30.5 to 30.9~mag (measured in a 0.2\arcsec\, radius circular aperture) for the NIRCam data in GOODS-S, highlighting that JADES is one of the deepest NIRCam surveys on the sky\footnote{Only matched in depth (in some bands) by the MIDIS/NIRCam-parallel project \citep{perez-gonzalez_life_2023} and The Next Generation Deep Extragalactic Exploratory Public Near-Infrared Slitless Survey (NGDEEP; \citealt{bagley_next_2024})}. On the other hand, we estimate a 5$\sigma$ depth ranging from 29.3 to 29.9~mag (measured in a 0.2\arcsec\, radius circular aperture) for the NIRCam data in GOODS-N. The total area covered by NIRCam in the GOODS fields is approximately 124 square arcmin\footnote{The JADES data can be downloaded from the following link: \url{https://archive.stsci.edu/hlsp/jades};}.

\subsubsection{HST}
For the HST data, we utilized ACS/WFC and WFC3/IR data from the Hubble Legacy Field  (HLF) observations that cover both fields, GOODS-N and GOODS-S. The HLF provides deep imaging in 9 HST bands covering a wide range of wavelengths (0.4$-$1.6$\mu$m), from the optical (ACS/WFC F435W, F606W, F775W, F814W, and F850LP filters) to the near-infrared (WFC3/IR F105W, F125W, F140W and F160W filters). We refer the reader to \citet{whitaker_hubble_2019} for a more detailed description of these observations\footnote{The HLF imaging is available at \url{https://archive.stsci.edu/ prepds/hlf/};}.

\subsubsection{NIRSpec}

We made use of the Near Infrared Spectrograph's Micro-Shutter Array (NIRSpec/MSA; \citealt{ferruit_near-infrared_2022, jakobsen_near-infrared_2022}) for the spectroscopic observations from the JADES NIRCam+NIRSpec program (PID: 1181, PI: Eisenstein), which cover the spectral range 0.6$–$5.3 $\mu$m, including observations with both the low-dispersion prism ($\text{R = 30–300}$) and all three medium-resolution gratings ($\text{R = 500–1500}$). We refer the reader to \citet{deugenio_jades_2024} for a more detailed description of this dataset.

\subsection{Sample Selection}
The ultra-deep NIRCam images from JADES in both GOODS-N and GOODS-S enable the photometric selection of red and compact sources, commonly referred to as LRDs, over a total area of $\approx 124$ arcmin$^2$. The extensive data, ranging from HST to NIRCam (the latter also offering medium bands), ensures robustness in photometric redshift estimation (see \citealt{deugenio_jades_2024, hainline_cosmos_2024}). To select our sample of LRDs in the GOODS fields, we made use of the public JADES DR2/DR3 catalogues. For both our selection and SED fitting, we made use of aperture photometry with $r = 0.25$\arcsec\; (i.e., \texttt{CIRC3}).

Since the initial identification of red and compact sources by \citet{labbe_population_2023}, significant efforts have been made to refine the photometric selection of LRDs. The selection criteria are reasonably effective in identifying broad-line (BL) AGNs. \citet{greene_uncover_2024} found that, of the sources followed up spectroscopically, approximately 60\%  (9/15) were confirmed as BL AGNs. However, 20\% of the candidate sources were brown dwarfs, indicating that an additional criterion against this type of contaminant is needed.

The spectra gathered so far of LRDs (e.g., \citealt{kocevski_hidden_2023, kokorev_uncover_2023, furtak_jwst_2023, greene_uncover_2024, matthee_little_2024}) reveal a defining feature: their SEDs appear blue at $1-2$ $\mu$m ($1000–2000$ \AA\; rest-frame) and red at 3-5 $\mu$m ($3100–5200$ \AA\; rest-frame), the so-called “v-shape”. 
With this in mind, and following the approach presented in \citet{kokorev_census_2024}, we adopt another color criterion (named {\tt brown dwarf removal}) that can potentially help in reducing the contamination from brown dwarfs, a result based on UNCOVER spectra (see \citealt{greene_uncover_2024}).

In addition, compared to the original selection, we  relaxed the criterion regarding adjacent filters to avoid excluding potential LRDs due to possible errors in the photometric measurements. Therefore, we visually inspected {\it all} selected sources and their SEDs\footnote{SEDs come from the official JADES catalogue, where \textsc{eazy} is adopted; see \citet{hainline_cosmos_2024} for more details.} to ensure the inclusion of genuinely red and compact objects, without misclassification due to strong emission lines.

\begin{figure*}[ht!]
    \centering
    \includegraphics[width = 0.98 \textwidth, height = 0.35 \textheight]{LRD_GOODS_sample.png}
    \caption{The photometrically selected LRD sample in the GOODS-N and GOODS-S fields at $z\approx4-8$, alongside other recent literature samples: \citet{labbe_population_2023, akins_cosmos-web_2024, barro_extremely_2024, kokorev_census_2024, perez-gonzalez_what_2024}. Our GOODS-S sample overlaps by 70\% with \citet{perez-gonzalez_what_2024} and by 75\% with \citet{kokorev_census_2024}. The differences arise because \citet{perez-gonzalez_what_2024} include sources above $z\approx 8$, which we did not consider in this work, and \citet{kokorev_census_2024} included sources for which we did not have coverage in both NIRCam/F115W and NIRCam/F200W, preventing us from fully applying our color criteria (see Section \ref{section2}).} 
    \label{fig:sample}
\end{figure*}

As demonstrated in various studies, using only color criteria can lead to the selection of red sources, whether they are true LRDs (i.e. red and compact) or red and extended. ``Compactness'' by itself is arbitrary. Therefore, following the example of recent works (e.g., \citealt{perez-gonzalez_what_2024, kokorev_census_2024}), we included a compactness criterion in the F444W band, which is the band that truly defines these sources as red. Specifically, we adopted the following criterion: $\text{F444W(0.5\arcsec)/F444W(0.25\arcsec)} < \text{1.7}$ (aperture diameters).

Thus, the criteria we adopted in this work are as follows: 
\begin{itemize} 
    \item \texttt{blue slope}: F150W $-$ F200W $<$ 0.8 mag 
    \item \texttt{red slope}: F277W $-$ F444W $>$ 0.7 mag 
    \item \texttt{brown dwarf removal}: F115W $-$ F200W $>$ $-0.5$ mag
    \item \texttt{compactness}: F444W(0.5\arcsec)/F444W(0.25\arcsec) $<$ 1.7 
\end{itemize}

We then restricted our selection to galaxies with photometric (or spectroscopic) redshifts within the range of $\approx 4-8$. This allowed us, for sources with $\text{F444W} \lesssim 28$~mag, to probe the UV part of the rest-frame spectrum of these sources with the SW bands from NIRCam (probing 0.2 to 0.4~$\mu$m at $z \approx 4$ and 0.1 to 0.2~$\mu$m at $z \approx 8$).

By applying the above color and compactness criteria, we initially selected $\approx 350$ candidates. After a thorough visual inspection of each source and its SED from \textsc{eazy} (see \citealt{hainline_cosmos_2024} for more details), we robustly identified 99 photometrically confirmed LRDs (see Figure \ref{fig:sample}). This includes 11 sources (marked with a cross) exhibiting LRD-like SEDs that were initially excluded for reasons such as strong emission lines (e.g., the complex H$\beta$ + [O III], which impact the F277W flux at $z\approx 5.5$ and thus the F277W $-$ F444W color), low SNR in the SW bands, incomplete filter coverage (e.g., GS206858, which is missing F150W and F200W), or faintness in F444W. These objects were included to account for their red and compact nature. In addition, 15 of our photometrically selected LRDs have NIRSpec spectra. Interestingly, \citet{zhang_abundant_2025} recently demonstrated that the selection criteria adopted in this work provide a more complete recovery of broad H$\alpha$ emitters with LRD-like SEDs from NIRCam/WFSS data, compared to the methods used in \citet{greene_uncover_2024} and \citet{barro_extremely_2024}, albeit with a trade-off in purity (see Figure 8 from \citealt{zhang_abundant_2025}).

\begin{figure*}[ht!]
    \centering
    \includegraphics[width=\textwidth]{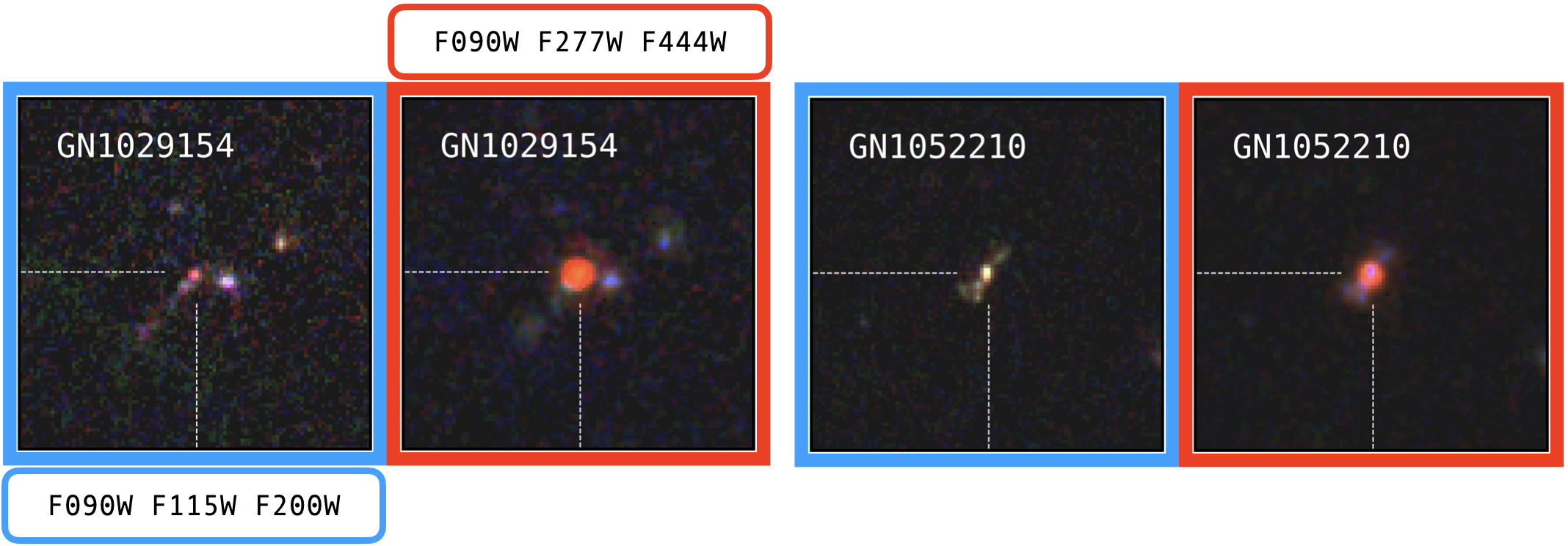}
    \caption{We display two examples from our sample of 99 photometrically selected LRDs in the GOODS-N and GOODS-S fields. For each source, we present two sets of $3\text{\arcsec}\times3\text{\arcsec}$ RGB postage stamps: one using the SW bands (F090W, F115W, and F200W) and another with the classic RGB colors (F090W, F277W, and F444W). A visual comparison between these two sets reveals that, for these two sources—which also have high SNR in the NIRCam SW bands—the morphology appears more complex at the shorter wavelengths compared to the classic compact morphology typically associated with LRDs at longer wavelengths (as highlighted in the classic RGB, in red), suggesting that \textit{they are not just a dot}.}
    \label{fig:mosaic_red_sample}
\end{figure*}

We show our selection in Figure \ref{fig:sample} and list their IDs\footnote{Throughout the paper, we refer to sources by their NIRCam IDs for simplicity.} and coordinates in Table \ref{tab:sources}. In Figure \ref{fig:mosaic_red_sample}, we present two examples of our selected LRDs where we show the comaparison between the NIRCam SW RGB (F090W, F115W, F200W) and the “classic” RGB (F090W, F277W, and F444W). This simple visual comparison reveals that LRDs can exhibit complex morphologies in the UV light, suggesting that \textit{they might not be just a dot}.

We cross-matched our sample with existing AGN catalogues in GOODS-S and GOODS-N. None of our sources appear in the pre-JWST AGN catalogues (\citealt{lyu_agn_2022}), which is expected given their limited coverage of high-redshift sources. However, six of our sources are included in the MIRI AGN catalogue presented by \citet{lyu_active_2024}. Among these, four have photometric redshifts consistent with those reported in \citet{lyu_active_2024}, while the remaining two have uncertain redshift estimates. Interestingly, four LRDs (GS197348, GN1001093, GN1061888, and GN1010816) were also identified in \citet{maiolino_jades_2023} and \citet{bunker_jades_2024}. The AGN classification carried out in \citet{lyu_active_2024} is primarily based on a significant MIRI photometric excess, consistent with hot dust emission from an obscured torus. While such a rising IR SED is atypical for the average LRD population, which often exhibits a flattening beyond $1\,\mu$m (e.g., \citealt{williams_galaxies_2024}), it is important to note that relying on average SEDs may overlook the intrinsic heterogeneity within this class. In this context, \citet{perez-gonzalez_what_2024} demonstrated that when MIRI data are included, photometrically selected LRDs display a broad range of behaviors, with some SEDs flattening and others rising, highlighting the diverse nature of the population. It is also worth mentioning two notable sources that follow the average LRD shape but exhibit a rising IR SED: the {\it Rosetta Stone} from \citet{juodzbalis_jades_2024} and {\it Virgil} from \citet{iani_midis_2024} and \citet{rinaldi_deciphering_2025}.

\begin{figure*}[ht!]
    \centering
    \includegraphics[width=\textwidth]{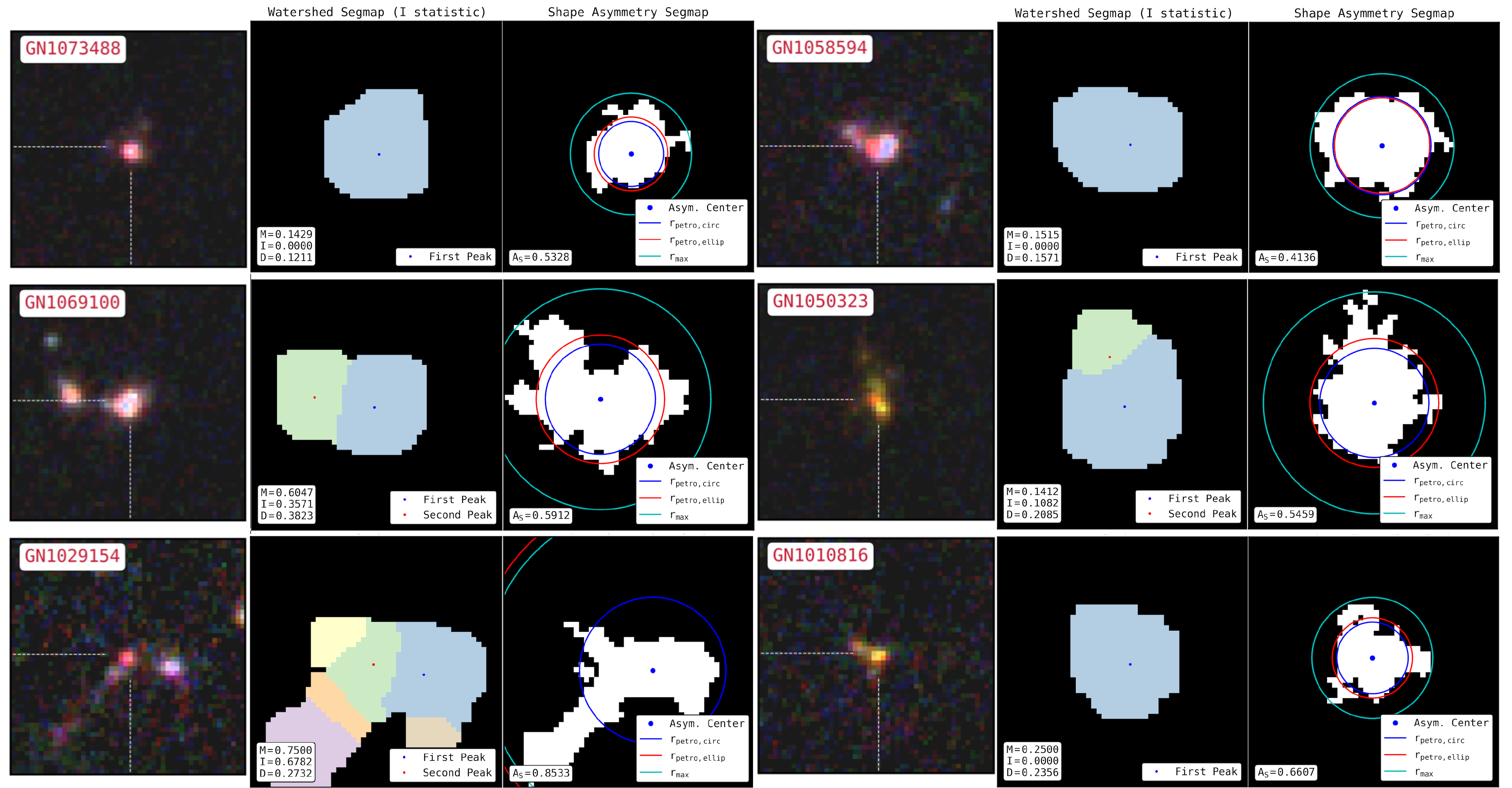}
    \caption{Example \textsc{statmorph} output images (middle and rightmost images in each image triplet) along with the associated SW RGB image of galaxy emission (leftmost image), showing the resulting \textit{MID} and \textit{$A_S$} values that comprise our quantitative LRD morphology study. The middle image of each source image triplet shows the segmentation map used specifically to calculate the \textit{I} statistic, where each distinct intensity maximum is highlighted with a different color. The sources with a single (blue) region represent single sources of emission, however 87\% of them show non-zero \textit{M} statistic values, indicating that the spatial footprints of multiple distinct source regions were detected (see Section 3 and the Appendix for a discussion of the \textit{M} and \textit{I} statistics). In the galaxy segmentation image used to calculate shape asymmetry, contained within the $r_{max}$ aperture (cyan line), $A_S$ values greater 0.2 mark a strong asymmetry/disturbance. It can be seen from the representative LRD examples shown here that they present highly asymmetric and complex morphologies, including systems with multiple sources, irregularly shaped single sources with multiple distinct regions of emission, and bright point-like sources embedded in fainter extended emission.} 

    \label{fig:lrd_statmorph_output}
\end{figure*}

\section{LRD Morphology}
\label{section3}

\subsection{Morphological analysis:  \textsc{statmorph}}

To study the UV morphology of the LRDs in our sample, we made use of the \textsc{statmorph} (\citealt{rodriguez-gomez_optical_2019}) software that computes a variety of morphology measures on an input image of a galaxy. To prepare image cutouts for input to the code, for each LRD we generated a 3\arcsec$\times$3\arcsec\; image with a pixel scale of 0.03"/px\footnote{For GOODS-S and GOODS-N, we adopted the following filters: F090W, F115W, F150W, F182M, F200W, and F210M. However, we caution that not all these bands were available for every source in our sample due to incomplete coverage in certain filters (e.g., in GOODS-S, F182M and F210M are only available from FRESCO and JEMS data).}, stacked over the PSF-matched\footnote{Before performing the morphological analysis with {\sc statmorph} on the stacked galaxy image cutouts, we PSF-matched all NIRCam SW bands to the reddest available band for each galaxy. Interestingly, a comparison of the {\sc statmorph} results before and after PSF-matching shows that the overall conclusions remain unchanged, further reinforcing the intrinsically complex UV morphology of these sources, as independently suggested in other studies (e.g., \citealt{chen_host_2025}).} SW filters to maximize SNR and ensure meaningful morphological measurements (a \textsc{statmorph} measure of  $>2.5$ SNR/pixel in the source aperture is considered to be trustworthy; refer to \citealt{rodriguez-gomez_optical_2019} for details). The image stacking also allowed us to identify the subset definitively showing extended structure and/or multiple apparently associated sources beyond the size of the SW FWHM. Out of 99 photometrically selected LRDs in GOODS-S and GOODS-N, we find that 30 LRDs in total display clear morphological features (both on the single SW images and the stacked SW bands), while the remaining 69 appear predominantly compact, lacking sufficient SNR for detailed characterization of any extended components, even when stacking the images. After verifying that the resulting subsample of 30 extended LRDs were detected with sufficient SNR/pixel value for analysis (in the range $4.8-12.8$), we computed their non-parametric morphological measurements. From among the suite of morphology indicators calculated by \textsc{statmorph}, we chose to utilize the non-parametric \textit{multimode-intensity-deviation} (\textit{MID}) statistics (\citealt{freeman_new_2013}) and \textit{shape asymmetry} (\textit{$A_S$}) parameter (\citealt{pawlik_shape_2016}). 

\subsubsection{Multimode-Intensity-Deviation (MID) and Shape Asymmetry ($A_S$) Statistics}

 The \textit{MID} statistics are useful for detecting multi-component systems, such as the double-nucleus of a late-stage merger; highly disordered post-merger remnants; galaxies with bright star-forming clumps in rest-UV emission; or an apparently single galaxy with an extended emission component(s). Briefly, the multimode ($M$) statistic identifies all non-contiguous groups of image pixels above a given intensity threshold and computes the area ratio of the top two largest regions. The intensity ($I$) statistic complements $M$ by calculating the intensity ratio between the two brightest regions in the galaxy image. Finally, the distance ($D$) statistic measures the normalized distance between the brightest local intensity maximum in the galaxy image and the centroid of the total emission, as identified in the binary detection mask (i.e., segmentation map). A more detailed explanation of these statistics can be found in the Appendix. The \textbf{$A_S$} parameter is a variation of the classic asymmetry parameter (\textit{A}) in that it is calculated using the binary detection mask as opposed to the flux image (\citealt{pawlik_shape_2016}). The \textit{$A_S$} parameter was designed in this way to detect the faint disturbances that appear along the edges of a galaxy merger remnant, such as wisps, cusps, and tidal tails, as well as the overall spatial asymmetry that characterizes an early- or late-stage merger as a singular system. In fact, \citet{nevin_accurate_2019} show \textit{$A_S$} to be the single most important non-parametric diagnostic of merger morphology in imaging data, over a maximal length of the merger lifetime.

\citet{freeman_new_2013} demonstrate that the \textit{MID} statistics in combination with the classic asymmetry parameter (\textit{A}) represent the most important set of non-parametric morphology indicators for accurately recovering known galaxy classifications\footnote{While the extensively studied \textit{concentration-asymmetry-smoothness} \textit{CAS} (\citealt{conselice_evidence_2003}) and \textit{Gini$-M_{20}$} (\textit{G$M_{20}$}) (\citealt{lotz_new_2004}) morphology diagnostics are also computed by \textsc{statmorph}, the reliability of these statistics is known to decrease along with decreasing galaxy size and SNR. Therefore, given the compact sizes and high redshift of LRDs, we utilize instead the (\textit{MID}) statistics that were introduced in \citet{freeman_new_2013} as an alternative to \textit{CAS} and \textit{G$M_{20}$} when such criteria hold.}, and do not show any systematic variation with galaxy size, degree of elongation, or SNR (at $\text{SNR}\gtrsim1.7$). However, given the unreliability of \textit{A} in instances of relatively low SNR and resolution (e.g. \citealt{conselice_evidence_2003}), as well as its inability to distinguish mergers from non-mergers over a significant fraction of the merger lifetime, we utilize the \textit{$A_S$} parameter instead, which outperforms \textit{A} in each of these instances (\citealt{nevin_accurate_2019}). 

The detailed algorithmic descriptions of these non-parametric morphology indicators are presented in the Appendix.

\begin{figure*}[ht!]
    \centering
    \includegraphics[width=\textwidth]{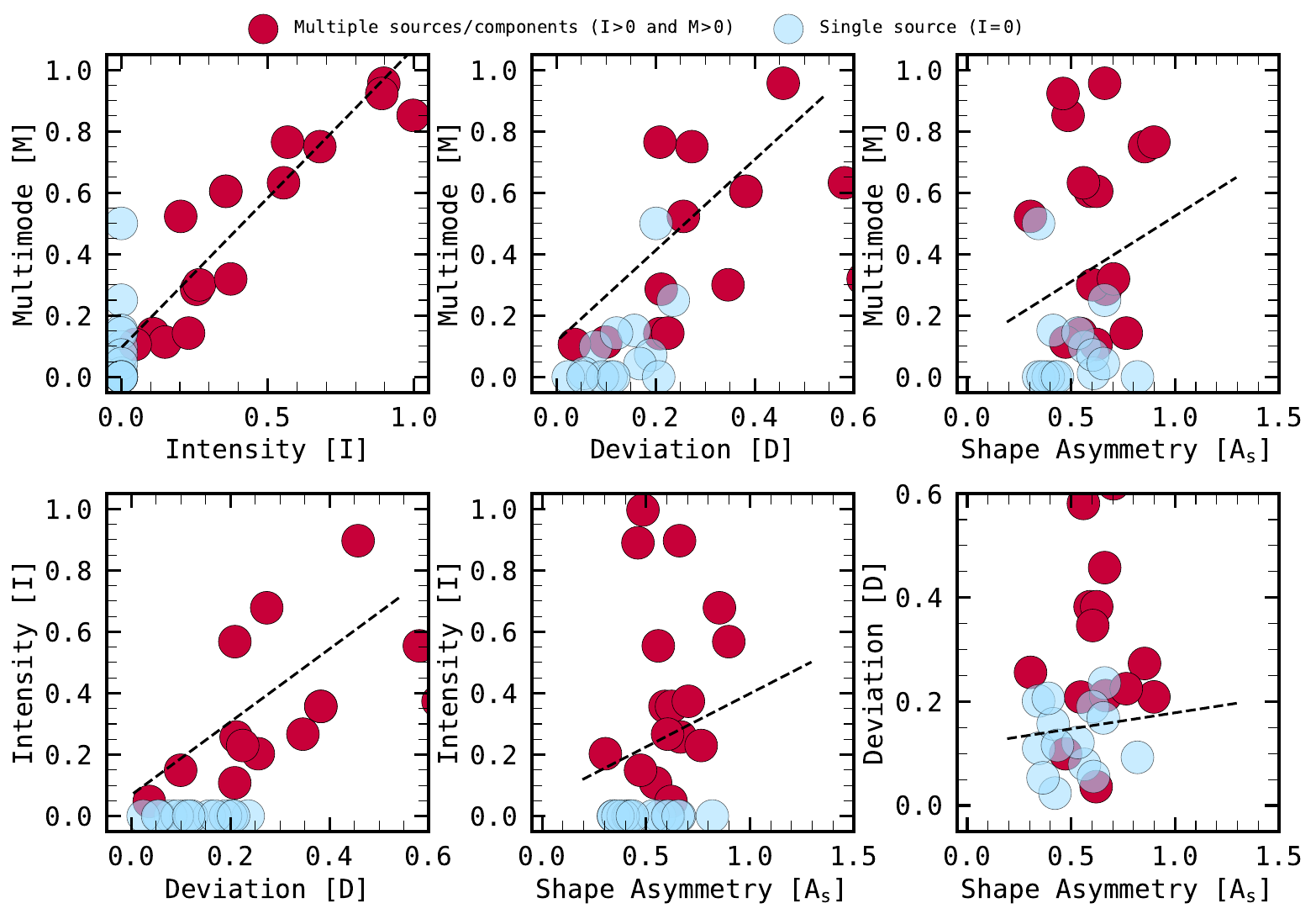}
    \caption{An adaptation of Figure 5 in \citet{freeman_new_2013}, where the \textit{MID} statistics for measuring galaxy morphologies are introduced, but with the classic \textit{asymmetry} parameter, \textit{A}, replaced by the \textit{shape asymmetry} parameter, $A_S$; see Section 3.1. The red symbols are akin to the merger candidates in the referenced work, as they represent the LRDs in our sample exhibiting multiple distinct emission and spatial components (indicated by non-zero \textit{I} and \textit{M} values, respectively). Close pairs of sources with similar brightnesses and sizes, such as major-merger candidates, would show \textit{I} and \textit{M} values tending towards a value of 1, while lower values of these two statistics would suggest a minor-merger candidate, or a single source with a relatively small and faint `companion' UV clump of emission. The blue symbols represent those LRDs identified with only a single (\textit{I} = 0), asymmetric ($A_S > 0.2$) source of emission, but with most ($87\%$) showing multiple non-contiguous pixel regions by their non-zero \textit{M} statistic values (similar to the “non-regular” non-merger candidates in \citealt{freeman_new_2013}). In short, all of the LRDs examined in the morphological analysis display irregular and extended features. The black lines represent the fit to the multi-source/multi-component LRDs (red points), demonstrating the expected positive linear relationship between these statistics. Based on the random forest regression and classification analysis of the combined \textit{MID} and \textit{A} statistics measured for 1639 galaxies in HST/WFC3 \textit{H}- and \textit{J}-band images in \citet{freeman_new_2013}, both the visually labeled (red) multi-component systems (classified as purely mergers in their study) and the (blue) non-merging irregular galaxies are detected with $\approx78\%$ accuracy (i.e., the percentage of correctly classified non-regular or merging galaxies).}
    \label{fig:lrd_morph_stats}
\end{figure*}

\subsection{The UV morphological properties of the LRDs}

In Figure~\ref{fig:lrd_statmorph_output}, we show the \textsc{statmorph} results of the \textit{MID} and \textit{$A_S$} statistic computation for a representative subset of the 30 LRDs in our sample appearing as multiple associated sources, irregularly shaped single sources, and apparent point sources embedded in fainter extended emission. In each triplet of images shown per source, the statistic values are displayed along with the corresponding segmentation map used in the calculation of each (\citealt{freeman_new_2013, pawlik_shape_2016, rodriguez-gomez_optical_2019}). Figure~\ref{fig:lrd_morph_stats} shows the measured relationship between \textit{$A_S$} and the \textit{MID} statistic values, which reveals that all LRDs included in the morphological analysis appear as strongly spatially disturbed systems, independent of the number of distinct sources or emission components detected, as indicated by \textit{$A_S$} values greater than 0.2 (\citealt{pawlik_shape_2016}). It can also be seen that \textit{$A_S$} is positively correlated with the corresponding non-zero (i.e., multi-component) \textit{M} and \textit{I} values, as expected where multiple components of emission are detected within/around the source (\citealt{freeman_new_2013}; see Appendix for details). Furthermore, the non-zero \textit{M} and \textit{I} statistic values are also positively correlated with one another, showing that the spatial area and brightness of a detected secondary source of emission tend to grow in tandem. The resulting values for the \textit{M} and \textit{I} statistics in our LRD sample show strong cases for multiple associated sources where their values tend towards 1, suggestive of a major merger, as well as candidates for minor mergers where the values are smaller (or one or more significant clumps of UV emission associated with a single source). 

For the $\approx50\%$ of cases with zero \textit{I} values (only one intensity peak identified in the galaxy emission) but non-zero \textit{M} values (multiple non-contiguous pixel groups identified above a given intensity threshold), it is clear from visual inspection that these represent LRDs with a single galaxy attached to an extended asymmetric emission structure. Finally, only in two cases do we find both zero \textit{I} and \textit{M} statistic values (i.e., a single and coincident spatial and emission component). However, both of these sources exhibit an asymmetric/disturbed spatial imprint, one with an \textit{$A_S$} value indicative of a mild spatial disturbance, and the other, strong. However, these two LRDs show particularly small sizes and relatively lower SNR compared to the rest of the sample examined in the morphological analysis, making it possible that they have multiple components that evaded detection by \textit{M} and \textit{I}.

Given that the \textit{D} statistic provides an independent measure of galaxy asymmetry (see Appendix) and therefore serves as a non-redundant complement to \textit{$A_S$}, we observe these two parameters to positively correlate, as expected, in instances where the LRD appears significantly extended, non-centralized, and disturbed. Furthermore, where a source happens to show a relatively symmetric spatial outline in the \textit{$A_S$} binary detection mask with equally weighted pixel values (tending towards lower values of \textit{$A_S$}), \textit{D} can still indicate a relatively high level of disturbance/disorder within the brightness distribution of the corresponding flux image of the galaxy, such as in the observed cases of an extended single source, but with its brightest peak of emission appearing off-center.

In Table \ref{tab:morph_params}, we summarize the main LRD properties estimated with \textsc{statmorph}. The \textit{MID} statistics indicate that 50\% of the LRDs selected for morphological analysis show at least two distinct, apparently associated sources or galaxy components, with the remainder appearing as single sources with highly asymmetric structure. We also notice from \textit{M} and \textit{I} that in $\approx50$\% of the multi-component LRDs, the two sources/regions used in the calculation of each statistic are of comparable brightness and size, suggesting they are potential major-merger candidates in the pre- or late-stage phase. To test whether the multiple components of the merger candidates are physically associated, we examined photometric redshifts and, where available, spectroscopic redshifts from NIRCam/WFSS observations (FRESCO and CONGRESS; \citealt{oesch_jwst_2023, lin_luminosity_2025}). For apparent companions with reliably extracted photometry, we find redshifts consistent with those of the LRDs, supporting a physical association rather than a chance projection. In cases where the companions are extremely close, reliable photometry is more difficult to obtain. However, aperture photometry within small radii often an SED shape consistent with that of the central LRD, further suggesting a potential physical connection, though firm conclusions remain uncertain.

\section{LRD Stellar Properties}
\label{section4}
\subsection{SED Fitting:  \textsc{bagpipes}}
We used \textsc{bagpipes} \citep{carnall_vandels_2019} to perform SED fitting and derive the stellar properties of the 99 photometrically selected LRDs. For the 15 galaxies with available spectroscopy, we also performed joint spectro-photometric fitting. In all cases, the redshift was fixed, based on either photometric or spectroscopic values.

\textsc{bagpipes} relies on synthetic templates from \citet{bruzual_stellar_2003} with a \citet{kroupa_variation_2001} IMF, adopting a cut-off mass of 100 $M_{\odot}$, and nebular emission modeled with \textsc{cloudy} \citep{ferland_2013_2013}. We employed a continuity non-parametric SFH model (\citealt{leja_how_2019}). For the latter, we defined the age bin edges (counted in look-back time from the redshift of the observation) based on each source's redshift (photometric or spectroscopic), following a logarithmic distribution from $z=30$ to the age of the Universe at that redshift. We adopted the same apporach for the age parameter. Stellar masses were allowed to range between $10^5$ and $10^{13}\ M_{\odot}$ (uniform prior in log). We opted for a Calzetti reddening law \citep{calzetti_dust_2000}, allowing $A_{V}$ to vary between 0 and 6, and allowed metallicity ($Z/Z_{\odot}$) to range from 0 to 2.5. The ionization parameter (U) was fixed at $-2$. Finally, each fit was calculated twice, with and without invoking an AGN component. We adopted the AGN implementation provided in {\sc bagpipes}, which models the continuum as a two-slope power law and includes broad Gaussian emission lines for H$\alpha$ and H$\beta$, following the approach of \citet{carnall_massive_2023} (see their Table 1). This model is sufficient to account for potential AGN contributions when estimating stellar properties, which is the main goal of this section. A detailed analysis of their AGN nature is presented in Section 5, based on the available NIRSpec spectra.

For the sources with NIRSpec spectra, we followed the techniques described by \citet{carnall_vandels_2019} to fully leverage the combination of spectroscopic and photometric data. As outlined in \citet{navarro-carrera_interstellar_2024}, we allowed for a $<2\sigma$ perturbation to the spectrum using a second-order Chebyshev polynomial to correct for systematic uncertainties in flux calibration. Additionally, we allow for a multiplicative factor on the spectroscopic errors to correct for underestimated uncertainties.

\begin{figure*}[ht!]
    \centering
    \includegraphics[width=\textwidth]{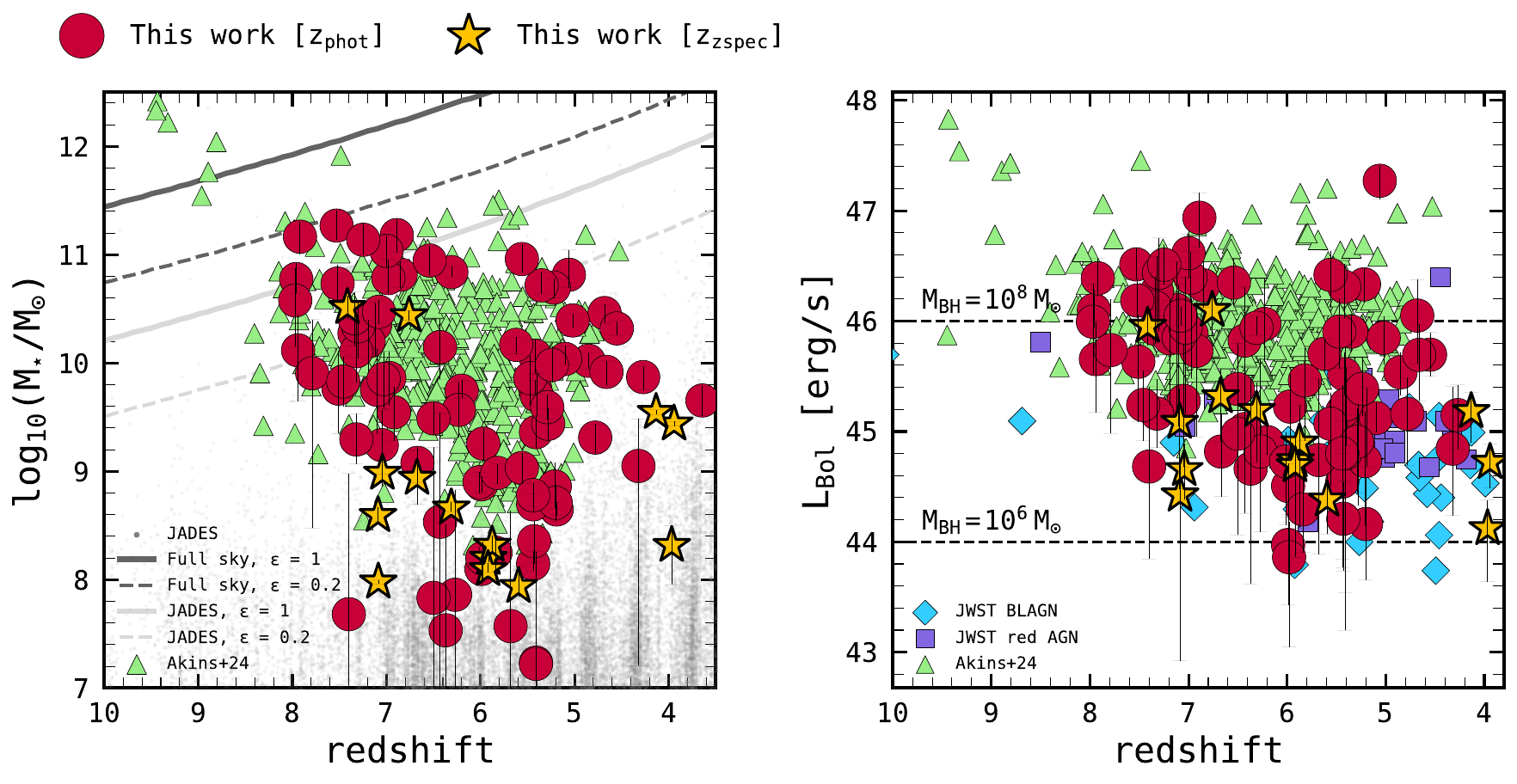}
    \caption{{\bf Left panel:} $M_{\star}$ as a function of redshift for our sample of 99 photometrically selected LRDs in the GOODS fields. In this plot, $M_{\star}$ comes from the {\sc bagpipes} runs with AGN. No significant differences arise when plotting, instead, $M_{\star}$ adopting stellar models only. Gray points represent galaxies from JADES DR2/DR3. The large sample of LRDs from \citet{akins_cosmos-web_2024} is presented for comparison. The sources with spectra are denoted by yellow stars. Grey points represent the JADES sources in both GOODS-S and GOODS-N. The allowed $M_{\star}$ as a function of redshift for two different star formation efficiencies values are also shown for the JADES area and the full sky. {\bf Right panel:} $L_{Bol}$ as a function of redshift. We computed $L_{Bol}$ from the intrinsic model SED (i.e., before any dust attenuation) by using the monochromatic luminosity at 5100 {\AA} and a bolometric correction of 9 (\citealt{richards_spectral_2006}). For comparison, we plot the LRD sample from \citet{akins_cosmos-web_2024} along with some other recent literature, divided into two groups: confirmed BL AGNs (\citealt{larson_ceers_2023, harikane_jwstnirspec_2023, maiolino_jades_2023, ubler_ga-nifs_2023, bogdan_evidence_2024, maiolino_small_2024, parlanti_ga-nifs_2024, ubler_ga-nifs_2024}) and red AGNs (\citealt{kokorev_uncover_2023, furtak_high_2024, greene_uncover_2024, matthee_little_2024}). Assuming an Eddington ratio = 1, we show what $L_{Bol}$ would correspond to log$_{10}(M_{BH}/M_{\odot}) = 6-8$ (horizontal dashed lines).}  
    \label{fig:lrds_prop}
\end{figure*}

\subsection{Results}

For runs allowing AGNs, we find that our LRDs have an average $A_{V} = 2.74^{+0.55}_{-0.71}$ mag  (16th and 84th percentile), consistent with the original results from \citet{labbe_population_2023}, where their sample of red, compact sources exhibited $A_{V} > 1.5$ mag. This also aligns with the recent findings from \citet{akins_cosmos-web_2024}, who identified a large population of red, compact objects in COSMOS. 
Our sources show, on average, log$_{10}(M_{\star}/M_{\odot})$ $=9.67_{-0.27}^{+0.17}$ (16th and 84th percentile), in agreement with the recent literature about LRDs (e.g., \citealt{akins_cosmos-web_2024}; see the left panel in Figure \ref{fig:lrds_prop}). On the other hand, the {\sc bagpipes} run using only stellar models indicates that our LRDs have, on average, $A_{V} = 1.16^{+0.11}_{-0.21}$ mag (16th and 84th percentile) and an average  log$_{10}(M_{\star}/M_{\odot})=9.07_{-0.08}^{+0.11}$ (16th and 84th percentile). To first order the lower $M_{\star}$ is a reflection of the lower extinction. These masses may, of course, be overestimated if the galaxies have IMFs that are more top-heavy than the Kroupa one used in {\sc bagpipes}.

Some sources in our sample exhibit very high $M_{\star}$, consistent also with the runs using stellar models alone. To interpret this, we estimate their implied star formation efficiency $\varepsilon$, defined as the fraction of baryons within a halo that are converted into stars, i.e., $\varepsilon = M_{\star} / (f_{\rm b}\, M_{\rm halo})$, where $f_{\rm b} = 0.156$ is the cosmic baryon fraction from \textit{Planck} and $M_{\rm halo}$ is the host halo mass inferred from abundance matching at fixed comoving number density (see also \citealt{boylan-kolchin_stress_2023}). We calculate the stellar mass ceiling corresponding to a maximal efficiency $\varepsilon=1$ and a more typical upper bound $\varepsilon=0.2$ by computing the cumulative halo mass function over the JADES survey area (GOODS-S $+$ GOODS-N)\footnote{ We made use of the python package {\tt hmf} and {\tt halomod} (\citealt{murray_hmfcalc_2013}) that can be found here: \url{https://github.com/halomod}. We adopted an SMT fitting function \citep{sheth_streaming_2001}.}.

\begin{figure}[ht!]
    \centering
    \includegraphics[width=1.0\linewidth]{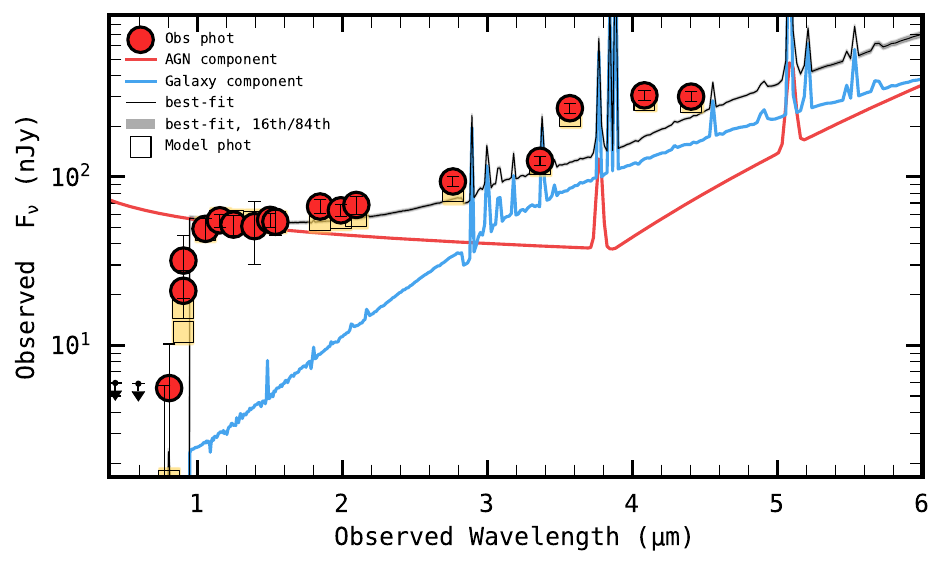}
    \caption{{\sc bagpipes} fit for JADES$-$GN$+$189.1520$+$62.2596 at $z_{\rm spec} = 6.759$. This object is also observed with NIRSpec medium-resolution gratings, which reveal broad components in both H$\beta$ and H$\alpha$ (Figure \ref{fig:BL_LRD}). The plot displays the best-fit galaxy and AGN templates from {\sc bagpipes}. This galaxy is part of the subsample exhibiting complex UV morphology (see Figure~\ref{fig:lrd_3_spec_examples}). The {\sc bagpipes} fit favors a galaxy template with high stellar mass and significant dust attenuation, consistent with findings from \citet{labbe_population_2023} and \citet{akins_cosmos-web_2024}, further illustrating the challenges involved in modeling the SEDs of such sources as also reported in \citet{rinaldi_deciphering_2025}.}  
    \label{fig:bagpipesp-fit}
\end{figure}

We find that a small number of galaxies in our sample lie above the $\varepsilon=1$ limit computed for the JADES area, similar to findings from the COSMOS-Web field by \citet{akins_cosmos-web_2024}. Two sources, which have NIRSpec spectra, show log${10}(M_{\star}/M_{\odot}) > 10$ at $z \approx 6.5–7.5$ and sit above $\varepsilon>0.2$. While one spectrum is of poor quality, the other (GN1010816) is robust and confirms a very high stellar mass of log${10}(M_{\star}/M_{\odot}) = 10.62^{+0.05}_{-0.05}$ at $z_{\rm spec} = 6.759$. Although such massive galaxies are expected to be rare at these redshifts, the implied star formation efficiencies challenge standard assumptions about baryon conversion and may point to alternative scenarios or underestimated halo masses. The {\sc bagpipes} fit for this object is shown in Figure~\ref{fig:bagpipesp-fit}, illustrating how the code models LRDs with high dust attenuation, as also found in \citet{akins_cosmos-web_2024}. In this fit, the AGN component contributes significantly to both the UV and optical portions of the spectrum. This once again highlights the challenges of SED fitting for such sources, as also demonstrated in a similar case at comparable redshift by \citet{rinaldi_deciphering_2025}. A more detailed analysis of these extreme sources, adopting different SED fitting codes, will be presented in a forthcoming study (Rinaldi et al. in prep).

We also estimated the bolometric luminosity ($L_{Bol}$) for our photometrically selected LRDs. We computed $L_{Bol}$ from the intrinsic best-fit SED (i.e., before any dust attenuation) using the monochromatic luminosity at 5100 {\AA} and applying a bolometric correction of 9 \citep{richards_spectral_2006}. In the right panel of Figure \ref{fig:lrds_prop}, we show our sample of LRDs along with recent literature results \citep{harikane_jwstnirspec_2023, maiolino_jades_2023, larson_ceers_2023, ubler_ga-nifs_2023, akins_cosmos-web_2024, bogdan_evidence_2024, furtak_high_2024, greene_uncover_2024, kokorev_census_2024, maiolino_small_2024, matthee_little_2024, parlanti_ga-nifs_2024, ubler_ga-nifs_2024}. We find that our sample agrees well with the region of parameter space covered by the large LRD sample presented in \citet{akins_cosmos-web_2024}, with a handful of sources reaching $L_{Bol}$ of up to $\approx 10^{47}\; \mathrm{erg/s}$.

\section{To be an AGN or not to be, that is the question: insights from NIRSpec}
\label{section5}
Some LRDs show broadening of the Balmer lines, suggesting  that they host AGNs (e.g., \citealt{greene_uncover_2024}).
To explore their properties, we follow  the approach outlined in \citet{kokorev_census_2024} to estimate the black hole mass ($M_{BH}$) under the assumption that our photometrically selected LRDs are {\it dominated} by AGNs. In general, a secure determination of $M_{BH}$ is not feasible; however, under this assumption, we can derive an estimate. In the recent literature, the Eddington rate ($\lambda_{Edd}$) for confirmed AGN in LRDs was found to range between 10\% and 40\% (e.g., \citealt{kokorev_uncover_2023, furtak_high_2024, greene_uncover_2024}). 

Following  the approach outlined in \citet{kokorev_census_2024}, we place a lower limit on the $M_{\text{BH}}$ by assuming that our AGN candidates accrete at the Eddington limit, i.e., $L_{bol} \approx L_{Edd}$ (where $L_{Edd}$ is directly proportional to $M_{BH}$). Under this assumption, the black-hole mass is given by
$M_{\rm BH} = L_{\rm bol} / (1.26 \times 10^{38}\;{\rm erg\,s^{-1}})\;M_\odot$, where the constant comes from the canonical expression for the Eddington luminosity (e.g., \citealt{rybicki_radiative_1979, peterson_introduction_1997}). Applying this conversion to our dust-corrected $L_{\rm bol}$ values yields a median $\log_{10}(M_{\rm BH}/M_\odot) \approx 7.40^{+0.30}_{-0.50}$ (16th and 84th percentile). If a significant fraction of the luminosity is derived from star formation, then this lower limit is overestimated. Using the scaling relation of \citealt{greene_intermediate-mass_2020} from the estimated median $M_{\star}$ yields an estimate an order of magnitude smaller.

These values are all consistent with the recently discovered population of red and compact sources (e.g., \citealt{akins_cosmos-web_2024, kokorev_census_2024}) and similar to the black hole masses in more traditional AGNs at similar redshift \citep{harikane_jwstnirspec_2023, maiolino_jades_2023}. That is, it is plausible that AGNs play an important role in any of our LRDs, including those without spectroscopic evidence for AGNs.

\begin{figure*}[ht!]
    \centering
    \includegraphics[width=\textwidth]{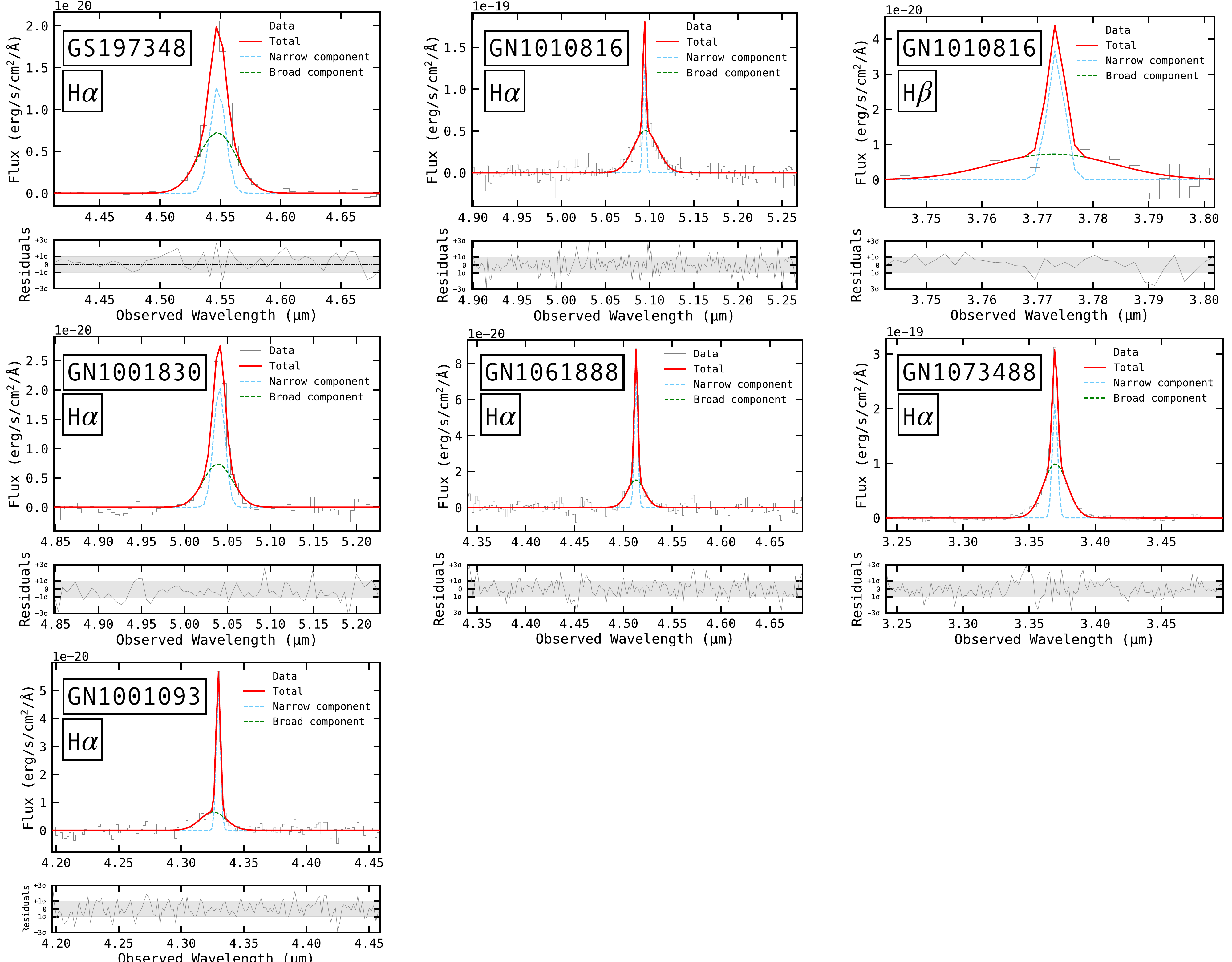}
    \caption{The 6 LRDs that show broadening in their H$\alpha$ from NIRSpec data (NIRCam IDs, see Table \ref{tab:sources}): GS197348, GN1010816, GN1001830, GN106188, and GN1073488. Particularly, GN1010816 (red frame) shows also broadening in its H$\beta$. Some of these objects are already reported in previous works, namely: GS197348, GN1010816, GN1001093, and GN1061888 (\citealt{maiolino_jades_2023, bunker_jades_2024}).}
    \label{fig:BL_LRD}
\end{figure*}

\begin{figure*}[ht!]
    \centering
    \includegraphics[width=\textwidth]{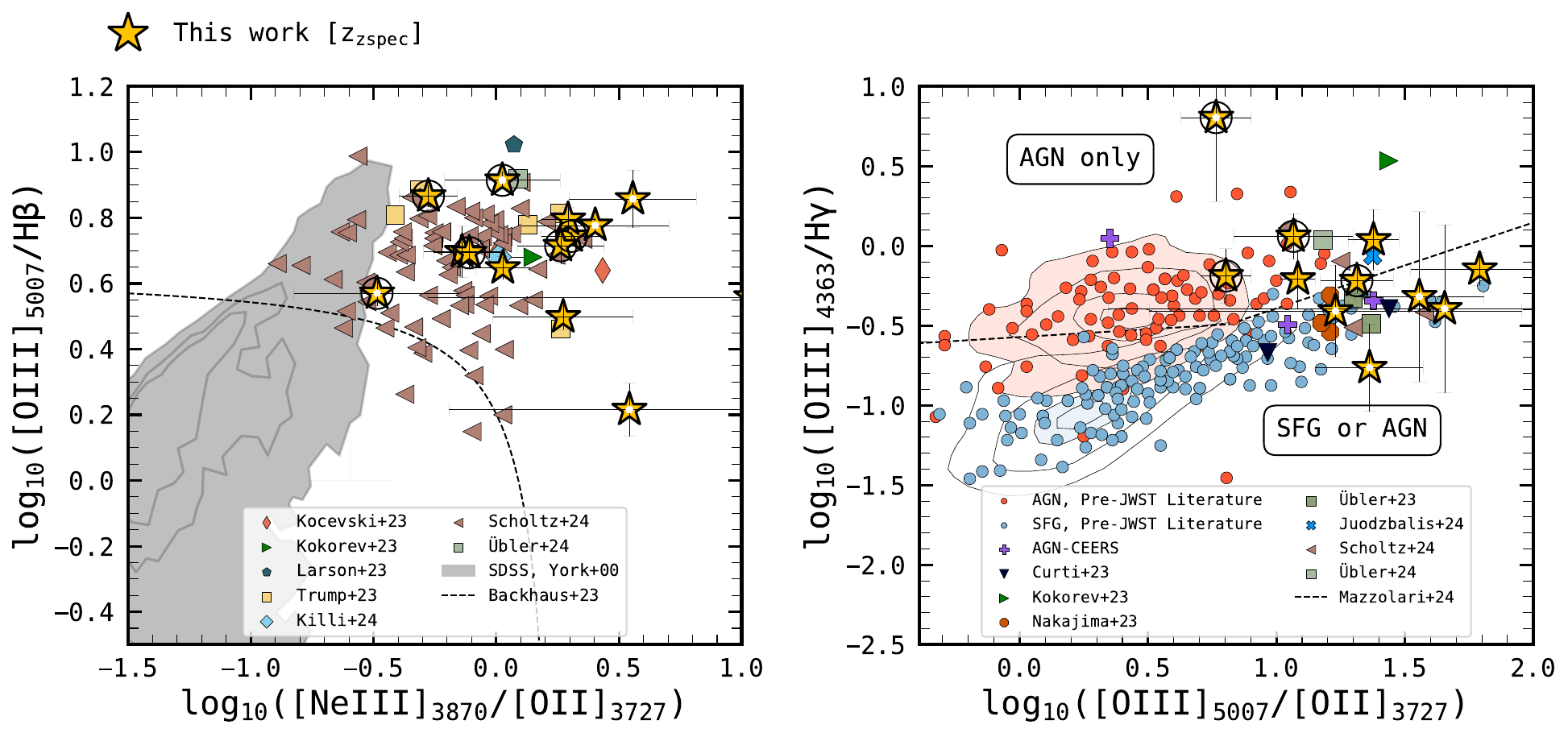}
    \caption{{\bf Left Panel:} We show the “OHNO” diagram. We present our sources as gold stars along with the recent literature from \citet{kocevski_hidden_2023, kokorev_uncover_2023, larson_ceers_2023, trump_physical_2023, killi_deciphering_2023, scholtz_jades_2023, ubler_ga-nifs_2024}. For comparison, we also show the SDSS sample (SFGs and AGNs) at low-z from \citet{york_sloan_2000}. The demarcation line comes from \citet{backhaus_clear_2023}.
    {\bf Right Panel:} We replicate the panel shown in \citet{mazzolari_new_2024}. Our sources are shown as gold stars along with some recent literature: e.g., \citet{curti_chemical_2023, kokorev_uncover_2023, nakajima_jwst_2023, ubler_ga-nifs_2023, juodzbalis_dormant_2024, scholtz_jades_2023, ubler_ga-nifs_2024}. We also show the demarcation line from \citet{mazzolari_new_2024}. Pre-JWST literature for AGNs and SFGs is shown as well (e.g., \citealt{izotov_broad-line_2007, amorin_extreme_2015}). The stars marked with black circles are the objects with enough SNR for morphological studies. The sources marked with a white dot are the ones with $\text{SNR}\lesssim2$ for one of the two line ratios involved in these diagnostics.}
    \label{fig:agn_sf_diagnostic}
\end{figure*}

Among our 99 photometrically selected LRDs in the GOODS fields, 15 have NIRSpec spectra, obtained with both the prism and the medium-resolution grating (\citealt{bunker_jades_2024, deugenio_jades_2024}), including 6 from our fiducial sample with sufficiently high SNR for morphological analysis. Therefore, we explored whether these sources exhibit spectral signatures that could indicate the presence of AGNs.
To do so, we made use of a modified version of the \textsc{MSAEXP} line-fitting algorithm (\citealt{brammer_msaexp_2023}), which allowed us to measure line fluxes, uncertainties, and observed equivalent widths for both the prism and gratings (Kokorev et al., in prep). We also implemented a custom routine into this code to fit individual (narrow and/or broad) emission lines, deblend them when necessary, and validate the results against \textsc{MSAEXP}. We then excluded one source from our analysis since the quality of its spectra (both prism and medium resolution gratings) prevented us from making proper estimation of any line fluxes. Finally, our findings were cross-matched with the official JADES line emission catalog, showing excellent agreement within the error bars.

Within our sub-sample of LRDs with NIRSpec spectra, we find that 6  (40\% of the spectroscopic sample) exhibit significant broadening in the H$\alpha$ emission line (either from prism or grating), with FWHM ranging from approximately 1200 km/s to 2900 km/s, as shown in Figure~\ref{fig:BL_LRD}, with 4 of them already presented in previous works (\citealt{maiolino_jades_2023, bunker_jades_2024}). Notably, one of these LRDs also shows broadening in the H$\beta$ line (observed with the medium-resolution grating), having FWHM $\approx 2000 \pm 500$ km/s. We caution the reader that the low resolution of PRISM may lead to significant instrumental broadening (e.g., \citealt{greene_uncover_2024}). Nevertheless, some sources relying solely on PRISM data, such as GS197348, have already been analyzed in previous studies (e.g., \citealt{bunker_jades_2023}). Three of the LRDs that show broad H$\alpha$ fall in our fiducial sample for morphological analysis. Unfortunately, the other 3 LRDs with broad H$\alpha$ lack sufficient SNR in the NIRCam SW bands (even by stacking the NIRCam SW bands), thus preventing  further study of their morphology. Interestingly, one of them, with SNR $\approx 4-6$ in the NIRCam SW channel (in the central region only), shows a faint component to the NW in F090W and F115W, which however is too faint to be observed in the stacked NIRCam SW image (F090W, F115W, F150W, F182M, F200W, and F210M).

To investigate further whether the selected LRDs with NIRSpec data could be classified as AGNs, we made use of two diagnostic plots based on the following emission lines: [OII]$\lambda \lambda$3727, 3728, [NeIII]$\lambda$3870, H$\beta$$\lambda$4861, and [OIII]$\lambda$5007. In particular, we explored the following ratios: [OIII]~$\lambda5007$/H$\beta$ vs. [NeIII]~$\lambda3870$/[OII]~$\lambda\lambda3727$ (commonly referred to as the “OHNO” diagram) and [OIII]~$\lambda4363$/H$\gamma$ vs. [OIII]~$\lambda5007$/[OII]~$\lambda\lambda3727$.
 Below, we present our results.

\paragraph{{\bf The OHNO diagram}} 
In the left panel of Figure \ref{fig:agn_sf_diagnostic}, we analyze our LRD sample by employing the “OHNO” diagram (\citealt{trouille_pushing_2011, zeimann_hubble_2015, backhaus_clear_2022, backhaus_clear_2023, cleri_using_2023, trump_physical_2023, feuillet_classifying_2024}), as the [NeIII]$\lambda$3870/[OII]$\lambda \lambda$3727, 3728 ratio has proven to be a robust ionization diagnostic for high-redshift galaxies (e.g., \citealt{backhaus_clear_2022, backhaus_clear_2023, backhaus_ceers_2024}). This diagnostic is particularly effective because [NeIII]$\lambda$3870 and [OII]$\lambda \lambda$3727, 3728 have similar ionization energy, and their being very close in wavelength minimizes the effects of dust attenuation. Particularly, it has been shown that employing this ratio can effectively help in distinguishing between star-forming galaxies (SFGs) and AGNs (see \citealt{zeimann_hubble_2015, backhaus_clear_2022, backhaus_clear_2023, backhaus_ceers_2024}). While this diagram is primarily sensitive to ionization, it also shows a dependence on metallicity (e.g., \citealt{tripodi_spatially_2024}). Therefore, we caution the reader that low-metallicity galaxies may introduce contamination, as already discussed in \citet{scholtz_jades_2023}.

On average, our photometrically selected LRDs exhibit $\mathrm{log_{10}([OIII]_{5007}/H\beta)} \gtrsim 0.5$ (median value of $0.71\pm0.03$), with the exception of one source that suffers from poor data quality in both prism and medium-resolution modes. Overall, the members of our sample show a consistently high [NeIII]$\lambda$3870/[OII]$\lambda \lambda$3727 ratio compared to the average SFG population. Interestingly, our sample occupies the same region of other LRDs recently studied in the literature (e.g., \citealt{ killi_deciphering_2023, kocevski_hidden_2023, kokorev_uncover_2023, larson_ceers_2023}) and, in general, the region occupied by selected broad and/or narrow line AGNs (e.g., \citealt{scholtz_jades_2023, ubler_ga-nifs_2024}). The gray area represents the sample of SFGs and AGNs selected from the Sloan Digital Sky Survey (SDSS) at $z\approx 0$ from \citet{york_sloan_2000}.

\paragraph{{\bf \boldmath{The [OIII]$\lambda$4363/H$\gamma$ vs. [OIII]$\lambda$5007/[OII]$\lambda$3727 diagram}}}
In the right panel of Figure \ref{fig:agn_sf_diagnostic}, we show another diagnostic diagram to investigate the nature of our photometrically selected LRDs: the [OIII]$\lambda$4363/H$\gamma$ vs. [OIII]$\lambda$5007/[OII]$\lambda$3727. This diagram has been recently presented in \citet{mazzolari_new_2024} and, as with the OHNO diagram, offers the advantage of using line ratios that lie very close in wavelength, therefore reducing the effects of dust reddening. 

Recent studies have demonstrated that some high-redshift galaxies ($z\gtrsim8$) exhibit anomalously high [OIII]$\lambda$4363 emission, potentially suggesting the presence of an AGN (e.g., \citealt{brinchmann_high-z_2023}). Intriguingly, this trend has been further confirmed by \citet{ubler_ga-nifs_2024}, who suggested that a strong [OIII]$\lambda$4363/H$\gamma$ ratio could actually result from higher interstellar medium (ISM) temperatures driven by AGN activity and, thus, point to the presence of an AGN. We remind the reader that, while this diagnostic is similar to the OHNO diagram, the separation between AGNs and SFGs in this case is primarily driven by differences in the gas temperature.

As demonstrated by \citet{mazzolari_new_2024}, normal SF galaxies (and their local analogs) tend to populate a well-defined region in this diagram, specifically the lower-right portion of the plot. In contrast, the AGN population spans a broader area, including the upper-left region. In particular, following the discussion presented by \citet{mazzolari_new_2024} who tested models computed by \citet{gutkin_modelling_2016}, \citet{feltre_nuclear_2016}, and \citet{nakajima_diagnostics_2022}, that area cannot be populated by any SFG model (see their Figure 1b for a more comprehensive view).  

The strength of [OIII]$\lambda$4363 emission line, which is produced through collisional excitation from high-energy levels, lies in its ability to provide key information about the gas temperature when compared to the H$\gamma$ intensity (e.g., \citealt{sanders_mosdef_2016}). This ratio can also offer insights into the metallicity of the ISM and the ionization parameter (e.g, \citealt{gburek_detection_2019}). The primary difference between SFGs and AGNs lies in the source of their ionizing radiation—young star clusters in the former and emission from the accretion disk in the latter—resulting in significantly more powerful ionizing radiation in the case of AGNs. This leads to higher electron temperatures, which, in turn, increase the [OIII]$\lambda$4363 emission and the [OIII]$\lambda$4363/H$\gamma$ ratio for a given ionization parameter (i.e., the ratio of the number density of incident ionizing photons and the number density of hydrogen atoms).

We observe that six sources lie above the separation line provided by \citet{mazzolari_new_2024}, and, more broadly, they overlap with the sample of AGNs (and some selected LRDs) identified in the recent studies (e.g., \citealt{curti_chemical_2023, kokorev_uncover_2023, nakajima_jwst_2023, scholtz_jades_2023, ubler_ga-nifs_2023, juodzbalis_dormant_2024, ubler_ga-nifs_2024}). Notably, one source is placed significantly above the separation line, although we caution the reader that the  spectral region of [OIII]$\lambda$4363/H$\gamma$ has low SNR for this source. Interestingly, this source does not exhibit any clear signature of broad components in its Balmer lines, but shows both [NII]$\lambda$6548 and [NII]$\lambda$6583 (detected with an SNR of approximately 4–6 in the medium-resolution grating). By inspecting its [NII]/H$\alpha$ ratio, we find that this source would lie precisely on the separation line between AGNs and SFGs in the classic “Baldwin, Phillips \& Terlevich” (BPT) diagram \citep{baldwin_classification_1981}.

More generally, our sample of photometrically selected LRDs is well constrained within a defined region of the parameter space, which overlaps with the area predominantly populated by recently discovered AGN using JWST. However, we note that this region, just below the demarcation line, also includes a subset of star-forming galaxies (pale blue circles; e.g., \citealt{izotov_broad-line_2007, amorin_extreme_2015, curti_chemical_2023, nakajima_jwst_2023, scholtz_jades_2023}).

\vspace{2.5mm}
Altogether, these two panels highlight the diversity of our LRD sample, ranging from pure AGNs to composite galaxies, underscoring their complex nature as a mix of different ionizing sources, as also suggested by previous studies (e.g., \citealt{perez-gonzalez_what_2024}). The consistency of our results, with our objects occupying a region of parameter space shared by AGN-dominated systems and star-forming galaxies, further supports their mixed nature. This suggests that AGN activity might significantly influence the properties of this subset of photometrically selected LRDs, though the extent of this influence remains unclear and warrants deeper spectroscopic investigation. We also note that these diagnostic plots are subject to uncertainties, including potential redshift dependencies (see discussion in \citealt{rinaldi_deciphering_2025}, Section 3.4.2), which must be considered when interpreting the nature of these sources. In Table~\ref{tab:emission_lines}, we report the line fluxes estimated for each source.

Finally, we computed $M_{BH}$ for the six sources that show broadening in H$\alpha$ and compared the estimated quantities with the recent literature. To estimate $M_{BH}$, we followed the approach presented in \citet{reines_dwarf_2013}, for which  $M_{BH}$ can be estimated as follows:

\begin{align}
\log_{10}\left(\frac{M_{\mathrm{BH}}}{M_{\odot}}\right) &= \alpha + \log_{10}(\varepsilon) + \beta \log_{10}\left(\frac{L_{H\alpha, \mathrm{broad}}}{1 \times 10^{42}\; \mathrm{erg/s}}\right) \nonumber \\
&\quad + \gamma \log_{10}\left(\frac{\mathrm{FWHM_{broad}}}{1 \times 10^{3}\; \mathrm{km/s}}\right),
\end{align}
where, $\alpha = 6.57$, $\beta = 0.47$, and $\gamma = 2.06$.

\begin{figure*}[ht!]
    \centering
    \includegraphics[width=\textwidth]{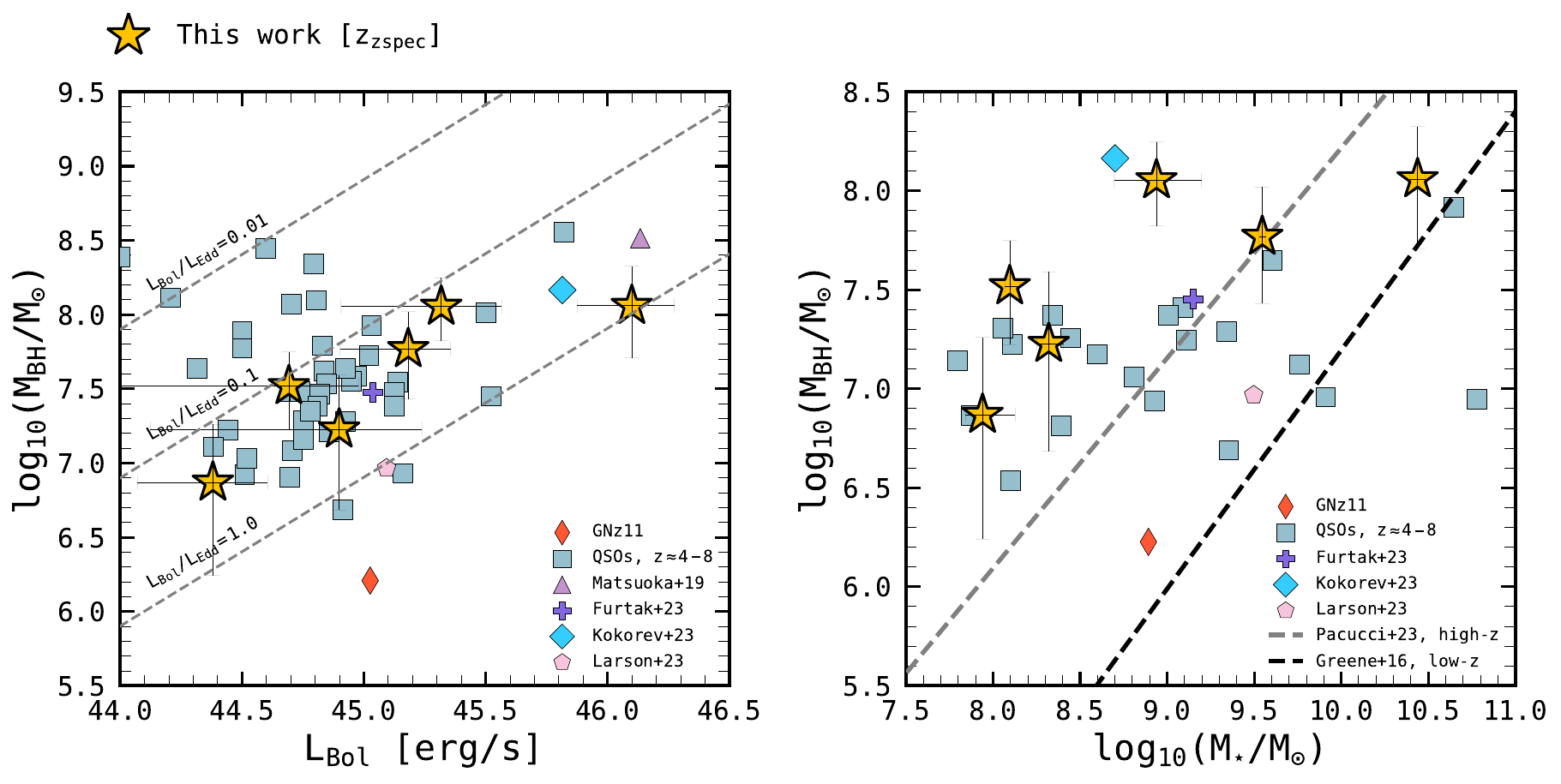}
    \caption{{\bf Left panel:} The derived black hole mass ($M_{\rm BH}$) as a function of bolometric luminosity ($L_{\rm Bol}$), in comparison with recent findings in the literature (e.g., GNz11, CEERS\_1019, the triply lensed quasar among others; \citealt{matsuoka_discovery_2019, furtak_jwst_2023, harikane_jwstnirspec_2023, kocevski_hidden_2023, kokorev_uncover_2023, larson_ceers_2023,  maiolino_jades_2023, maiolino_small_2024, matthee_little_2024}). The dashed lines represent bolometric luminosities corresponding to Eddington ratios of $L_{\rm Bol}/L_{\rm Edd}$ = 0.01, 0.1, and 1.0. {\bf Right panel:} The black hole-to-stellar mass relation is presented, with the bold black dashed line indicating the best fit to $z \approx 0$ AGN samples (\citealt{greene_megamaser_2016}). The trend at higher redshifts is based on the recent analysis by \citet{pacucci_jwst_2023}. Color codes and markers are the same as for the left panel.}
    \label{fig:lrd_prop_bh}
\end{figure*}

The results are shown in Figure \ref{fig:lrd_prop_bh}. Overall, our findings align with recent studies of selected LRDs analyzed using spectroscopic data (e.g., \citealt{kokorev_uncover_2023}). They also tend to validate our rough estimates made previously.

\section{Morphology of LRDs with Broad H$\alpha$}
\label{section6}
\begin{figure*}[ht!]
    \centering
    \includegraphics[width=\textwidth]{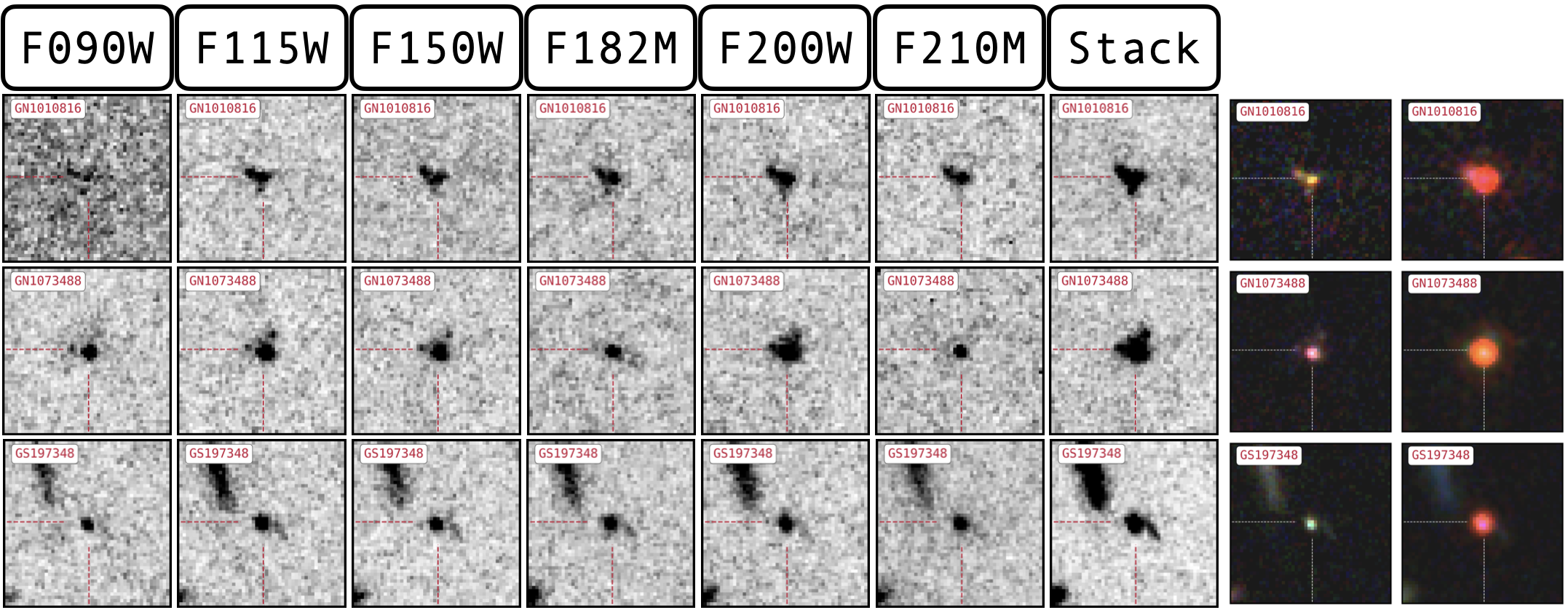}
    \caption{The LRDs in our sample with NIRSpec spectra that show broad H$\alpha$ line emission (and with sufficient SNR for morphological characterization), possibly indicating the presence of an AGN. Postage stamps ($1.5$\arcsec $\times$ $1.5$\arcsec) from the NIRCam SW channel (F090W, F115W, F150W, F182M, F200W, and F210M). For visual comparison, we also show the RGB postage stamps considering the SW bands only and the “classic” RGB (i.e., F090W, F277W, and F444W). }
    \label{fig:lrd_3_spec_examples}
\end{figure*}

Among the LRD candidates with NIRSpec spectra, three of them show a sufficiently high SNR and extent for morphological analysis, as well as broad H$\alpha$ (Figure ~\ref{fig:lrd_3_spec_examples}). Given the potential for both AGN and star-forming activity that these features imply (in combination with the results of the various spectral line-ratio diagnostics discussed in the previous section), we consider that their highly disturbed appearances may be associated with one or both of these processes, and potentially result from merging activity. As shown in Figure ~\ref{fig:lrd_3_spec_examples}, GN1010816 is detected with multiple distinct regions of emission (four, according to its \textit{I} statistic measure), at least two of which are of comparable intensity and size; and with a total spatial extent measured to have significant asymmetry. Given the sampling of rest-UV emission in the stacked NIRCam SW-channel images used for morphological characterization, this example could indicate multiple merging galaxies, triggering bursts of star formation in the form of UV clumps (e.g., \citealt{guo_clumpy_2015}), and/or one or two unobscured AGN. It is also possible that the multiple components of emission do not represent distinct nuclei in the act of merging, but a single galaxy (possibly in the post-merger phase) with a surrounding clumpy structure from merger- and/or AGN-triggered star formation, and/or clumpy accretion onto an AGN (e.g., \citealt{degraf_black_2017}). 

The other two LRDs with broad H$\alpha$ line emission appear as single sources with a highly asymmetric spatial footprint, as determined through both human and computer vision. GN1073488 appears distinctly bright and point-like and is  embedded in an asymmetric diffuse structure. GS197348, on other hand, appears elongated, with a relatively long, narrow, and faint emission structure appearing to “shoot off” from one of its sides. This faint extended feature is notably missed by the \textit{MID} statistics, but is caught by the \textit{$A_S$} algorithm, which was explicitly designed to detect such faint edge features. If this feature is real and associated with the LRD - which it appears to be based on its presence in all the NIRCam SW filters considered -  it could potentially be a manifestation of AGN feedback, e.g. the UV emission that has been found to spatially coincide with AGN radio jets (e.g., \citealt{rubinur_study_2024}).

\section{Summary and Discussion}
\label{section7}

In this study, we analyzed a sample of 99 photometrically identified LRDs in the GOODS fields, selected using color and compactness criteria (Figure \ref{fig:sample} and \ref{fig:mosaic_red_sample}; e.g., \citealt{labbe_population_2023}).

We examined the rest-wavelength UV morphology of these LRDs by analyzing ultra-deep NIRCam SW images using the \textsc{statmorph} software. Out of 99 photometrically selected LRDs, 30\% of them show extended structure and also with sufficient SNR in these bands to allow for a meaningful morphological analysis. The remaining 70\% are strongly dominated by sources $\lesssim$ 400 pc in diameter and lack extended components even in stacked SW band images. We found that all these objects exhibit $A_{S} > 0.2$, with a median value of $\approx 0.5$, suggesting that these sources are generally highly spatially disturbed and likely to be undergoing mergers or interactions. Such elevated $A_{S}$ values align with recent findings (e.g., \citealt{bonaventura_relation_2024}), which suggest that galaxies with $A_{S} > 0.2$ are frequently linked to ongoing or recent merger activity (Figure \ref{fig:lrd_statmorph_output} and  \ref{fig:lrd_morph_stats}). 

While most studies consider the rest-optical emission of galaxies in diagnosing merger morphologies, we have analyzed the rest-UV morphology of LRDs at $z\approx4-8$ imaged in the NIRCam SW bands. The expectation from the results of \citet{mager_galaxy_2018} is that the UV asymmetry of merging/peculiar galaxies will be consistent with, or appear more pronounced than, that measured in the rest-optical, for all galaxy types: these authors  observe a significant increase in the \textit{clumpiness} of all galaxy types at shorter wavelengths.  Furthermore, more recent studies on high-z galaxy morphology have shown that, in general, galaxies do not exhibit dramatic changes when transitioning from UV to optical light. For instance, \citet{treu_early_2023} studied a sample of Lyman Break Galaxies during the Epoch of Reionization and found that, within the uncertainties and scatter, the classic morphological indices ($G$, $M_{20}$, \textit{A}, etc.) remain relatively consistent across different wavelengths. 
That is, the measures of galaxy morphology in our study should reflect those expected from the rest-optical emission. However, a caveat to consider in the rest-UV is that, without knowledge of the quantity, distribution, optical depth, and covering fraction of the dust present in a galaxy, the extent to which it may be affecting the observed rest-UV morphology cannot be constrained.  While it is possible that the presence of severely attenuating dust could contribute to a clumpy UV morphology by blocking from view all but the brightest UV emission regions, it is unlikely that the 30 star-forming LRDs included in our morphological analysis appear with such irregular and extended features purely as a result of this effect.

The disturbed LRD morphology we observe could be driven by gravitational interactions that channel gas toward the central regions, potentially fueling both star formation and black hole growth, finally leading to an AGN phase. However, quite surprisingly, preliminary analysis of the stacked X-ray emission of the LRDs using the deepest available \textit{Chandra} ACIS-I imaging coverage, from the CDFS 7Ms survey, shows no detection; this mirrors the findings of \citet{yue_stacking_2024}, who were equally baffled by the non-detection of stacked X-ray emission of their LRD sample. Gas accretion onto an AGN should exhibit significant X-ray luminosity, as well as the shock-heated gas of a major, gas-rich galaxy merger (e.g., \citealt{cox_x-ray_2006} and references therein). Therefore, it is conceivable that, in X-ray-undetected LRDs, a dynamical or other physical process is preventing the angular momentum of the gas in the system from dropping low enough to funnel onto the nucleus, causing a delay to the infall of tidal material and the consequent detectable X-ray emission. For instance, several galaxy merger simulations suggest that the higher gas turbulence and velocity dispersion observed in high-redshift galaxies in comparison to their lower-redshift counterparts inhibits the propagation of gas inflows towards the center of the system, possibly resulting in suppressed AGN activity in gaseous, high-redshift galaxies (see \citealt{shah_investigating_2020} and references therein). In any case, low X-ray luminosity has been found for a variety of other high-redshift AGNs \citep{lyu_active_2024, maiolino_jwst_2024}, so the puzzle is not confined to LRDs. 

To put the morphology results in context, we employed {\sc bagpipes} to analyze the  properties of all 99 LRDs and found that, on average, the sources exhibit $A_{V} \approx 2.74^{+0.55}_{-0.71}$  (16th and 84th percentile) mag when AGN models are included, and $A_{V} \approx 1.16_{-0.21}^{+0.11}$ mag (16th and 84th percentile) when using stellar models only. The average stellar mass is log$_{10}(M_{\star}/M_{\odot}) \approx 9.67^{+0.17}_{-0.27}$ (16th and 84th percentile), or  log$_{10}(M_{\star}/M_{\odot})=9.07_{-0.08}^{+0.11}$  (16th and 84th percentile) when stellar models only are considered. To first order, the difference in $M_{\star}$ is a reflection of different dust extinctions.
We also estimated their (dust-corrected) $L_{Bol}$ from their best-fit spectra (assuming a correction factor of $\approx 9$, \citealt{richards_spectral_2006}). 
By making the assumption that all these selected LRDs host AGNs and adopting an Eddington ratio of 1, we derived lower limits for their $M_{BH}$, with a median value of log$_{10}(M_{BH}/M_{\odot}) \approx 7.40^{+0.30}_{-0.50}$ {\bf (16th and 84th percentile)}, or an order of magnitude smaller if we estimate BH masses from the usual relation with the $M_{\star}$ (\citealt{greene_intermediate-mass_2020}). These results are also consistent with those reported for the recently discovered population of red and compact sources (e.g., \citealt{akins_cosmos-web_2024, kokorev_census_2024}), i.e., our sample is typical of the general class. 

Among our sample of LRDs, 15 have NIRSpec spectra, which have been explored to investigate whether they host an AGN. We employed three different diagnostic diagrams to evaluate the state of their ISM. We find a variety of behavior, ranging from those classified as pure AGNs to those showing a mixed nature (i.e., classified as composite galaxies), indicating their complex nature (Figure \ref{fig:agn_sf_diagnostic}). Interestingly, six of them exhibit broadening in their H$\alpha$ lines, with one also showing broadening in H$\beta$. The remaining 60\% show no clear signs of AGN presence (i.e., no broadening in their Balmer lines); nonetheless, the diagnostic plots employed in this study hint at either AGN activity or a mixed nature for these sources, suggesting that deeper spectroscopic data are needed to further investigate their nature. For those showing broad H$\alpha$, we estimated their $M_{BH}$ from the broad H$\alpha$ component (\citealt{reines_dwarf_2013}) and found that our results are consistent with recent findings about LRDs (Figure \ref{fig:lrd_prop_bh}; e.g., \citealt{furtak_jwst_2023, kokorev_uncover_2023, larson_ceers_2023}).

A significant portion of our LRD sample exhibits disturbed UV morphology, with some objects clearly observed in a merger state (Figure \ref{fig:lrd_3_spec_examples}).  
Two sources with NIRSpec data show very high $M_{\star}$ and highly disturbed UV morphology (Figure \ref{fig:lrds_prop}). The most reliably modeled of these sources shows a very high $M_{\star}$ (log$_{10}(M_{\star}/M_{\odot})=10.62_{-0.05}^{+0.05}$) at $z_{spec} = 6.759$.

The mechanisms that trigger rapid gas accretion onto super massive black holes (SMBHs) remain still unclear, which directly ties into the nature of the LRDs in our sample. A compelling theoretical hypothesis is that galaxy mergers and interactions drive AGN activity by funneling gas into the central regions of galaxies, thereby fueling the SMBH (e.g., \citealt{gabor_comparison_2016, blumenthal_go_2018}), also recently supported by \citet{duan_galaxy_2024}. However, observational evidence remains inconclusive, with several studies finding no definitive correlation (e.g., \citealt{villforth_morphologies_2014, hewlett_redshift_2017, ellison_definitive_2019, kocevski_ceers_2023, pierce_galaxy_2023}). Given that our morphological analysis revealed significant asymmetries and signs of disturbance in a substantial fraction of our LRDs, it is plausible that interactions might play a substantial role in triggering AGNs in these systems (although we do not have a control sample to put this on a more quantitative basis). This would be an important, although perhaps not unexpected, difference from the situation at lower redshift. Further spectroscopic and morphological studies, particularly those utilizing deep NIRSpec/IFU data, will be essential in unveiling the true nature of the LRDs and exploring the connection between their disturbed UV morphologies and potential AGN activity.

\tabletypesize{\scriptsize}
\begin{deluxetable*}{llccc|llccc}
\tablecaption{List of the photometrically selected LRDs in the GOODS fields \label{tab:sources}}
\tablewidth{0pt}
\tablehead{
\colhead{JADES ID} & \colhead{NIRCam ID} & \colhead{Redshift} & \colhead{R.A. (deg)} & \colhead{Dec. (deg)} &
\colhead{JADES ID} & \colhead{NIRCam ID} & \colhead{Redshift} & \colhead{R.A. (deg)} & \colhead{Dec. (deg)}
}
\startdata
JADES-GN+189.1797+62.2246 & 1001093 & 5.595* & 189.1797 & 62.2246 & JADES-GS+53.0763-27.9099 & 2532 & 6.98 & 53.0763 & -27.9099 \\
JADES-GN+189.0915+62.2281 & 1001830 & 6.675* & 189.0915 & 62.2281 & JADES-GS+53.1191-27.8926 & 11786 & 7.34 & 53.1191 & -27.8926 \\
JADES-GN+189.1096+62.2285 & 1001895 & 7.52   & 189.1096 & 62.2285 & JADES-GS+53.1072-27.8906 & 13418 & 6.50 & 53.1072 & -27.8906 \\
JADES-GN+189.1277+62.2326 & 1002836 & 7.10   & 189.1277 & 62.2326 & JADES-GS+53.1136-27.8848 & 19348 & 5.41 & 53.1136 & -27.8848 \\
JADES-GN+189.0963+62.2391 & 1004685 & 7.414* & 189.0963 & 62.2391 & JADES-GS+53.0570-27.8744 & 35453 & 5.10 & 53.0570 & -27.8744 \\
JADES-GN+189.1517+62.2594 & 1010767 & 6.20   & 189.1517 & 62.2594 & JADES-GS+53.0641-27.8709 & 39376 & 7.00 & 53.0641 & -27.8709 \\
JADES-GN+189.1520+62.2596 & 1010816 & 6.759* & 189.1520 & 62.2596 & JADES-GS+53.0558-27.8690 & 41769 & 7.79 & 53.0558 & -27.8690 \\
JADES-GN+189.2038+62.2684 & 1013041 & 7.089* & 189.2038 & 62.2684 & JADES-GS+53.1304-27.8607 & 54648 & 6.37 & 53.1304 & -27.8607 \\
JADES-GN+189.0571+62.2689 & 1013188 & 7.32   & 189.0571 & 62.2689 & JADES-GS+53.1153-27.8592 & 57356 & 4.27 & 53.1153 & -27.8592 \\
JADES-GN+189.0385+62.2693 & 1013282 & 7.12   & 189.0385 & 62.2693 & JADES-GS+53.1083-27.8510 & 70714 & 6.50 & 53.1083 & -27.8510 \\
JADES-GN+189.0659+62.2733 & 1014361 & 4.32   & 189.0659 & 62.2733 & JADES-GS+53.0605-27.8484 & 73690 & 5.40 & 53.0605 & -27.8484 \\
JADES-GN+189.0721+62.2734 & 1014406 & 5.19   & 189.0721 & 62.2734 & JADES-GS+53.1476-27.8420 & 79803 & 5.41 & 53.1476 & -27.8420 \\
JADES-GN+189.0506+62.2794 & 1016275 & 7.97   & 189.0506 & 62.2794 & JADES-GS+53.1127-27.8383 & 82737 & 5.21 & 53.1127 & -27.8383 \\
JADES-GN+189.0577+62.2836 & 1017514 & 5.02   & 189.0577 & 62.2836 & JADES-GS+53.0732-27.8331 & 86916 & 7.05 & 53.0732 & -27.8331 \\
JADES-GN+189.0612+62.2841 & 1017694 & 7.34   & 189.0612 & 62.2841 & JADES-GS+53.1281-27.8292 & 89635 & 5.99 & 53.1281 & -27.8292 \\
JADES-GN+188.9878+62.2911 & 1020140 & 4.66   & 188.9878 & 62.2911 & JADES-GS+53.1338-27.8283 & 90354 & 7.96 & 53.1338 & -27.8283 \\
JADES-GN+189.1131+62.2924 & 1020485 & 5.26   & 189.1131 & 62.2924 & JADES-GS+53.1590-27.8183 & 99267 & 6.67 & 53.1590 & -27.8183 \\
JADES-GN+189.1590+62.2602 & 1029154 & 5.62   & 189.1590 & 62.2602 & JADES-GS+53.1593-27.8117 & 104238 & 5.28 & 53.1593 & -27.8117 \\
JADES-GN+189.0409+62.2693 & 1030265 & 5.42   & 189.0409 & 62.2693 & JADES-GS+53.1019-27.8109 & 104849 & 5.24 & 53.1019 & -27.8109 \\
JADES-GN+189.1798+62.2824 & 1032447 & 7.086* & 189.1798 & 62.2824 & JADES-GS+53.1408-27.8022 & 110739 & 5.916* & 53.1408 & -27.8022 \\
JADES-GN+189.0870+62.2908 & 1033797 & 5.22   & 189.0870 & 62.2908 & JADES-GS+53.1254-27.7874 & 120484 & 7.08 & 53.1254 & -27.7874 \\
JADES-GN+189.1983+62.2970 & 1034762 & 7.043* & 189.1983 & 62.2970 & JADES-GS+53.1269-27.7862 & 121710 & 7.92 & 53.1269 & -27.7862 \\
JADES-GN+189.2586+62.1432 & 1037138 & 7.51   & 189.2586 & 62.1432 & JADES-GS+53.1728-27.7831 & 124327 & 7.94 & 53.1728 & -27.7831 \\
JADES-GN+189.2395+62.1444 & 1037341 & 5.68   & 189.2395 & 62.1444 & JADES-GS+53.2040-27.7721 & 132229 & 7.247* & 53.2040 & -27.7721 \\
JADES-GN+189.2346+62.1475 & 1037974 & 7.46   & 189.2346 & 62.1475 & JADES-GS+53.1908-27.7679 & 136872 & 7.19 & 53.1908 & -27.7679 \\
JADES-GN+189.2707+62.1484 & 1038147 & 5.82   & 189.2707 & 62.1484 & JADES-GS+53.1479-27.7599 & 143133 & 6.43 & 53.1479 & -27.7599 \\
JADES-GN+189.2062+62.1505 & 1038673 & 6.43   & 189.2062 & 62.1505 & JADES-GS+53.1582-27.7391 & 154428 & 6.54 & 53.1582 & -27.7391 \\
JADES-GN+189.2631+62.1512 & 1038849 & 3.91   & 189.2631 & 62.1512 & JADES-GS+53.0789-27.8842 & 165902 & 5.56 & 53.0789 & -27.8842 \\
JADES-GN+189.2940+62.1531 & 1039353 & 5.29   & 189.2940 & 62.1531 & JADES-GS+53.0877-27.8712 & 172975 & 4.78 & 53.0877 & -27.8712 \\
JADES-GN+189.2436+62.1549 & 1039805 & 5.26   & 189.2436 & 62.1549 & JADES-GS+53.0557-27.8688 & 174121 & 7.30 & 53.0557 & -27.8688 \\
JADES-GN+189.2024+62.1627 & 1042541 & 5.41   & 189.2024 & 62.1627 & JADES-GS+53.0374-27.8656 & 175930 & 5.35 & 53.0374 & -27.8656 \\
JADES-GN+189.3216+62.1627 & 1042550 & 7.45   & 189.3216 & 62.1627 & JADES-GS+53.0964-27.8531 & 184838 & 7.32 & 53.0964 & -27.8531 \\
JADES-GN+189.2735+62.1665 & 1043804 & 5.84   & 189.2735 & 62.1665 & JADES-GS+53.1060-27.8482 & 187025 & 6.92 & 53.1060 & -27.8482 \\
JADES-GN+189.3395+62.1848 & 1050323 & 6.89   & 189.3395 & 62.1848 & JADES-GS+53.1265-27.8181 & 197348 & 5.919* & 53.1265 & -27.8181 \\
JADES-GN+189.1748+62.1901 & 1052210 & 6.01   & 189.1748 & 62.1901 & JADES-GS+53.0677-27.8123 & 198980 & 4.68 & 53.0677 & -27.8123 \\
JADES-GN+189.1493+62.2075 & 1058594 & 3.64   & 189.1493 & 62.2075 & JADES-GS+53.1548-27.8065 & 200576 & 6.31 & 53.1548 & -27.8065 \\
JADES-GN+189.1680+62.2170 & 1061888 & 5.874* & 189.1680 & 62.2170 & JADES-GS+53.1214-27.7949 & 203749 & 7.53 & 53.1214 & -27.7949 \\
JADES-GN+189.2248+62.2258 & 1064405 & 5.20   & 189.2248 & 62.2258 & JADES-GS+53.1135-27.7935 & 204022 & 7.45 & 53.1135 & -27.7935 \\
JADES-GN+189.2613+62.2320 & 1065744 & 5.56   & 189.2613 & 62.2320 & JADES-GS+53.1386-27.7903 & 204851 & 5.42 & 53.1386 & -27.7903 \\
JADES-GN+189.2292+62.1462 & 1068797 & 5.06   & 189.2292 & 62.1462 & JADES-GS+53.1390-27.7844 & 206858 & 3.941* & 53.1390 & -27.7844 \\
JADES-GN+189.2141+62.1490 & 1069100 & 5.44   & 189.2141 & 62.1490 & JADES-GS+53.1661-27.7720 & 210600 & 6.310* & 53.1661 & -27.7720 \\
JADES-GN+189.2793+62.1501 & 1069299 & 5.47   & 189.2793 & 62.1501 & JADES-GS+53.1792-27.7587 & 214552 & 5.97 & 53.1792 & -27.7587 \\
JADES-GN+189.2358+62.1681 & 1072112 & 5.43   & 189.2358 & 62.1681 & JADES-GS+53.1925-27.7531 & 216165 & 5.99 & 53.1925 & -27.7531 \\
JADES-GN+189.1974+62.1772 & 1073488 & 4.132* & 189.1974 & 62.1772 & JADES-GS+53.1848-27.7440 & 217926 & 6.97 & 53.1848 & -27.7440 \\
JADES-GN+189.3075+62.1780 & 1073625 & 6.23   & 189.3075 & 62.1780 & JADES-GS+53.1583-27.7409 & 218515 & 5.98 & 53.1583 & -27.7409 \\
JADES-GN+189.1786+62.1872 & 1075363 & 5.44   & 189.1786 & 62.1872 & JADES-GS+53.1614-27.7377 & 219000 & 6.85 & 53.1614 & -27.7377 \\
JADES-GN+189.1493+62.2083 & 1079572 & 3.966* & 189.1493 & 62.2083 & JADES-GS+53.1248-27.8663 & 283663 & 4.55 & 53.1248 & -27.8663 \\
JADES-GN+189.2816+62.2161 & 1081040 & 4.85   & 189.2816 & 62.2161 & --- & --- & --- & --- & --- \\
JADES-GN+189.2854+62.2235 & 1081928 & 6.27   & 189.2854 & 62.2235 & --- & --- & --- & --- & --- \\
JADES-GN+189.0962+62.2392 & 1113205 & 7.40   & 189.0962 & 62.2392 & --- & --- & --- & --- & --- \\
JADES-GN+189.2143+62.1491 & 1119051 & 5.44   & 189.2143 & 62.1491 & --- & --- & --- & --- & --- \\
JADES-GN+189.1364+62.2226 & 1177425 & 7.03   & 189.1364 & 62.2226 & --- & --- & --- & --- & --- \\
\enddata
\tablecomments{Sources with spectroscopic redshifts ($z_{\mathrm{spec}}$) are marked with * (\citealt{bunker_jades_2023, deugenio_jades_2024}). Photometric redshifts come from \textsc{eazy} (\citealt{hainline_cosmos_2024}).}
\end{deluxetable*}

\begin{deluxetable*}{ccccccccc}
\tablecaption{Morphological parameters for the LRDs with enough SNR for morphological analysis\label{tab:morph_params}}
\tablewidth{0pt}
\tablehead{
\colhead{NIRCam ID} & \colhead{$r_{\mathrm{max,circ}}$} & \colhead{SNR/pixel} & \colhead{$A_s$} & \colhead{$M$} & \colhead{$I$} & \colhead{$D$}
}
\startdata
11786   & 6.315 & 3.483 & 0.661 & 0.957 & 0.897 & 0.458 \\
13418   & 4.736 & 6.932 & 0.620 & 0.106 & 0.048 & 0.036 \\
19348   & 3.127 & 6.876 & 0.390 & 0.000 & 0.000 & 0.206 \\
79803   & 5.376 & 10.515 & 0.461 & 0.923 & 0.890 & 0.631 \\
172975  & 4.164 & 13.419 & 0.604 & 0.071 & 0.000 & 0.190 \\
174121  & 6.434 & 9.504 & 0.604 & 0.300 & 0.267 & 0.346 \\
175930  & 5.230 & 44.287 & 0.819 & 0.000 & 0.000 & 0.093 \\
187025  & 4.982 & 14.451 & 0.655 & 0.043 & 0.000 & 0.168 \\
197348  & 14.825 & 12.124 & 0.704 & 0.319 & 0.374 & 0.619 \\
204851  & 14.683 & 8.208 & 0.485 & 0.852 & 0.997 & 0.776 \\
206858  & 14.244 & 13.344 & 0.559 & 0.632 & 0.554 & 0.581 \\
210600  & 5.292 & 15.948 & 0.365 & 0.000 & 0.000 & 0.053 \\
214552  & 8.968 & 6.201 & 0.765 & 0.143 & 0.230 & 0.225 \\
219000  & 3.726 & 24.490 & 0.434 & 0.000 & 0.000 & 0.116 \\
1010816 & 5.758 & 6.466 & 0.661 & 0.250 & 0.000 & 0.236 \\
1017514 & 4.615 & 9.338 & 0.343 & 0.500 & 0.000 & 0.201 \\
1020140 & 4.578 & 13.954 & 0.344 & 0.000 & 0.000 & 0.110 \\
1029154 & 14.055 & 5.502 & 0.853 & 0.750 & 0.678 & 0.273 \\
1033797 & 4.364 & 4.478 & 0.564 & 0.098 & 0.000 & 0.079 \\
1038147 & 4.549 & 22.171 & 0.608 & 0.012 & 0.000 & 0.056 \\
1050323 & 8.081 & 10.632 & 0.546 & 0.141 & 0.108 & 0.209 \\
1052210 & 8.159 & 15.250 & 0.474 & 0.114 & 0.149 & 0.099 \\
1058594 & 7.587 & 22.272 & 0.414 & 0.152 & 0.000 & 0.157 \\
1065744 & 4.562 & 15.248 & 0.422 & 0.000 & 0.000 & 0.024 \\
1069100 & 9.045 & 27.571 & 0.591 & 0.605 & 0.357 & 0.382 \\
1069299 & 23.252 & 7.181 & 0.666 & 0.286 & 0.258 & 0.211 \\
1073488 & 4.978 & 18.402 & 0.533 & 0.143 & 0.000 & 0.121 \\
1079572 & 22.575 & 3.778 & 0.898 & 0.765 & 0.568 & 0.209 \\
1081040 & 15.603 & 8.041 & 0.304 & 0.523 & 0.203 & 0.256 \\
1119051 & 9.138 & 27.570 & 0.623 & 0.605 & 0.357 & 0.382 \\
\enddata

\tablecomments{This table lists morphological parameters extracted from \textsc{starmorph} for the LRD with sufficient SNR for robust morphological analysis. Columns include the maximum circular radius $r_{\mathrm{max,circ}}$, SNR per pixel, shape asymmetry ($A_s$), multiplicity ($M$), intensity ($I$), and deviation ($D$).}
\end{deluxetable*}

\tabletypesize{\scriptsize}
\begin{deluxetable*}{ccccccc}
\tablecaption{Emission Line Measurements for LRDs with NIRSpec data \label{tab:emission_lines}}
\tablehead{
\colhead{NIRCam ID} & \colhead{[O III]$\lambda$5007} & \colhead{H$\beta$} & \colhead{[Ne III]$\lambda$3870} & \colhead{[O II]$\lambda$3727} & \colhead{H$\gamma$} & \colhead{[O III]$\lambda$4363} \\
\colhead{} & \colhead{($10^{-20}$ erg/s/cm$^2$)} & \colhead{($10^{-20}$ erg/s/cm$^2$)} & \colhead{($10^{-20}$ erg/s/cm$^2$)} & \colhead{($10^{-20}$ erg/s/cm$^2$)} & \colhead{($10^{-20}$ erg/s/cm$^2$)} & \colhead{($10^{-20}$ erg/s/cm$^2$)}
}
\startdata
197348 & $454.93 \pm 9.19$ & $82.06 \pm 1.71$ & $33.63 \pm 2.19$ & $16.60 \pm 4.60$ & --- & --- \\
110739 & $744.32 \pm 11.30$ & $167.73 \pm 13.23$ & $65.33 \pm 13.37$ & $61.43 \pm 9.53$ & $82.37 \pm 15.76$ & $51.02 \pm 19.41$ \\
206858 & $594.88 \pm 27.17$ & $160.08 \pm 11.82$ & $33.40 \pm 23.90$ & $102.21 \pm 31.84$ & $2.84 \pm 1.86^*$ & $18.15 \pm 18.53^*$ \\
210600 & $723.67 \pm 41.44$ & $98.51 \pm 4.85$ & $34.09 \pm 6.94$ & $64.45 \pm 11.85$ & --- & --- \\
1010816 & $1200.75 \pm 22.27$ & $146.24 \pm 8.46$ & $109.16 \pm 9.90$ & $103.10 \pm 55.01^*$ & $59.46 \pm 11.19$ & $68.35 \pm 18.46$ \\
1001093 & $226.02 \pm 8.77$ & $62.44 \pm 11.59$ & $54.70 \pm 17.70$ & $4.99 \pm 15.09^*$ & $17.22 \pm 9.77^*$ & $6.95 \pm 7.45^*$ \\
1001830 & $531.57 \pm 14.61$ & $73.80 \pm 14.83$ & $53.12 \pm 7.66$ & $14.75 \pm 8.54$ & $21.89 \pm 8.22$ & $10.52 \pm 12.32^*$ \\
1004685 & $84.27 \pm 7.88$ & $51.27 \pm 8.29$ & $17.30 \pm 4.76$ & $4.96 \pm 8.28^*$ & $19.58 \pm 5.00$ & $7.63 \pm 4.67^*$ \\
1013041 & $918.91 \pm 18.09$ & $153.56 \pm 7.42$ & $37.59 \pm 11.85$ & $14.85 \pm 9.16^*$ & $67.78 \pm 8.45$ & $48.49 \pm 9.70$ \\
1032447 & $1231.22 \pm 20.99$ & $197.28 \pm 16.30$ & $101.14 \pm 10.06$ & $51.55 \pm 11.74$ & $55.17 \pm 18.15$ & $60.27 \pm 17.16$ \\
1034762 & $298.76 \pm 11.13$ & $60.08 \pm 10.63$ & $35.04 \pm 7.15$ & $48.28 \pm 14.29$ & $38.36 \pm 14.20$ & --- \\
1061888 & $283.32 \pm 7.26$ & $89.99 \pm 7.77$ & $23.03 \pm 10.33$ & $12.29 \pm 5.91$ & $109.07 \pm 17.45$ & $18.80 \pm 11.50^*$ \\
1073488 & $1161.28 \pm 14.23$ & $223.88 \pm 28.73$ & $103.35 \pm 25.05$ & $56.44 \pm 18.47$ & $145.62 \pm 12.02$ & $88.34 \pm 14.60$ \\
1079572 & $680.03 \pm 18.51$ & $137.64 \pm 15.40$ & $83.01 \pm 20.43$ & $106.93 \pm 23.94$ & $109.61 \pm 15.76$ & $71.11 \pm 25.83$ \\
\enddata
\tablecomments{Emission line measurements are given as the flux $\pm$ error in units of $10^{-20}$ erg/s/cm$^2$. Sources with $^*$ have $\text{SNR} < 2$ for the respective line.}
\end{deluxetable*}

\acknowledgments

The authors thank an anonymous referee for a careful reading and useful comments on this manuscript.

This work is based on observations made with the NASA/ESA/CSA JWST. The data were obtained from the Mikulski Archive for Space Telescopes at the Space Telescope Science Institute, which is operated by the Association of Universities for Research in Astronomy, Inc., under NASA contract NAS 5-03127 for JWST. These observations are associated with JWST programs GTO \#1180, GO \#1210, GO \#1963, GO \#1895, and \# 3215. The authors acknowledge the FRESCO, JEMS, and \# 3215 teams led by coPIs P. Oesch, C. C. Williams, M. Maseda, D. Eisenstein, and R. Maiolino for developing their observing program with a zero-exclusive-access period. Processing for the JADES NIRCam data release was performed on the lux cluster at the University of California, Santa Cruz, funded by NSF MRI grant AST 1828315. Also based on observations made with the NASA/ESA Hubble Space Telescope obtained from the Space Telescope Science Institute, which is operated by the Association of Universities for Research in Astronomy, Inc., under NASA contract NAS 526555.

KIC acknowledges funding from the Dutch Research Council
(NWO) through the award of the Vici Grant
VI.C.212.036.

ST acknowledges support by the Royal Society Research Grant G125142.

H\"U acknowledges support through the ERC Starting Grant 101164796 ``APEX''.

PGP-G acknowledges support from grant PID2022-139567NB-I00 funded by Spanish Ministerio de Ciencia e Innovaci\'on MCIN/AEI/10.13039/501100011033, FEDER, UE.

RM acknowledges support by the Science and Technology Facilities Council (STFC), by the ERC through Advanced Grant 695671 “QUENCH”, and by the UKRI Frontier Research grant RISEandFALL. RM also acknowledges funding from a research professorship from the Royal Society.

SC acknowledges support by European Union’s HE ERC Starting Grant No. 101040227 - WINGS.

WMB acknowledges support by a research grant (VIL54489) from VILLUM FONDEN.

ECL acknowledges support of an STFC Webb Fellowship (ST/W001438/1).

BER acknowledges support from the NIRCam Science Team contract to the University of Arizona, NAS5-02015, and JWST Program 3215.

The research of CCW is supported by NOIRLab, which is managed by the Association of Universities for Research in Astronomy (AURA) under a cooperative agreement with the National Science Foundation.

AJB acknowledges funding
from the “FirstGalaxies” Advanced Grant from the European Research Council (ERC) under the European Union’s Horizon 2020
research and innovation program (Grant agreement No. 789056).

MR, CNAW, BDJ, and EE acknowledge the JWST/NIRCam contract to the University of Arizona NAS5-02015.

\clearpage

\section{Appendix A}

Here we present the mathematical formalisms underpinning the \textit{Multimode-Intensity-Deviation} (\textit{MID}) and \textit{shape asymmetry} \textit{$A_S$} morphology indicators chosen to characterize the LRDs of the present study. We utilized the \textsc{statmorph} code for the morphology analysis, which reads a galaxy image and its associated weight map (the 1 $\sigma$ error image, also known as the ‘sigma image’ in Galfit and similar image analyses) to examine both the brightness and spatial distribution of its pixel values. It then calculates various galaxy morphology metrics, including the following non-parametric measures of galaxy structure: asymmetry (A), clumpiness (S), concentration index (C), Gini index (G), and moment of light (M20) (Lotz et al. 2004, Wu 1999, Bershady et al. 2000, Conselice et al. 2000.  \textsc{statmorph} computes the \textit{MID} and \textit{$A_S$} statistics as detailed below, with slight modifications to their original definitions as presented in Freeman et al. (2013) and Pawlik et al. (2016), respectively (\citealt{rodriguez-gomez_optical_2019}).

\section{\texttt{Multimode-Intensity-Deviation (MID) Statistics}}

 The \textit{statmorph} software calculates the non-parametric \textit{Multimode-Intensity-Deviation} (\textit{MID}) statistics according to their original definitions in Freeman et al. (2013), except for the \textbf{Multimode (M)} statistic, for which a modified version defined in Peth et al. (2016) is adopted. The algorithm defining the \textit{M} statistic locates all the non-contiguous groups of image pixels that lie above a given intensity threshold, \textit{q}, sorts them by pixel area, then computes the area ratio of the second-largest group ($A_{q,2}$) to the first ($A_{q,1}$); this process is then repeated for a number of intensity thresholds. \textit{M} is represented by the maximum area ratio resulting from all the trials: 
 
 \begin{align}
     \textbf{\textit{M} = $max_q$($A_{q,2}$/$A_{q,1}$)} 
 \end{align}
 
 As such, \textit{M} values approaching 1 are likely to indicate a double-nucleus, while values close to zero should be interpreted as a single source (i.e., where the non-zero detection of a relatively small, secondary pixel group is likely to be noise). 

 The \textbf{Intensity (I)} statistic serves as a complement to \textit{M} in that it computes the intensity ratio of the two brightest source regions in the galaxy image. It does so by first locating all the distinct intensity maxima in a smoothed image of the galaxy emission; identifying and summing the intensities of the group of pixels belonging to each intensity peak; and then calculating the ratio of the summed intensity of the second-brightest pixel group ($I_2$) to the first ($I_1$): 
 
 \begin{align}
     \textbf{ \textit{I} = $I_2$/$I_1$}
 \end{align}

 It should be noted that, due to the different ways in which the \textit{M} and \textit{I} statistics are calculated, they may not simultaneously detect multiple sources/components of emission in a given galaxy image. In other words, \textit{M} is a function of the spatial footprint of distinct regions in the galaxy emission, while \textit{I} considers their relative intensities. However, this can be a useful difference in cases of relatively faint emission components/sources and/or low resolution, where one of the statistics can provide an independent, compensating measure where the other statistic fails to detect multiple image components (or confirms the finding of the other where both agree). Such a scenario could occur where \textit{M} detects multiple distinct regions of emission within the galaxy image, but \textit{I} does not resolve more than one local intensity maximum. There can also be a case where the \textit{I} statistic locates a relatively small but bright secondary region of emission, that leads to a disproportionately small value of M. 

The \textbf{Deviation (D)} statistic provides a measure of the normalized distance between the centroid of the total extent of the galaxy emission as identified in the segmentation map, and its brightest local intensity maximum. This statistic is therefore useful in identifying irregular/peculiar galaxy shapes -- e.g. late-stage or post-coalescent mergers, or active sources experiencing significant spatial disruption from star-forming and/or AGN processes -- given the expectation that symmetric and ordered morphologies such as spheroids and disks would show a \textit{D} value close to zero. It is calculated according to the following formula, where the image centroid is represented by ($x_{c},y_{c}$), the brightest peak resulting from the \textit{I} statistic calculation is ($x_{I_{1}}$,$y_{I_{1}}$), and $n_{seg}$ is the number of pixels in the segmentation map.

\begin{align}
    \textbf{D = $\sqrt{\pi/n_{seg}}$ $\sqrt{(x_{c}-x_{I_{1}})^2 + (y_{c}-y_{I_{1}})^2}$}
\end{align}

\section{\texttt{Shape Asymmetry ($A_S$)}}

The \textbf{$A_S$} parameter is calculated in exactly the same way as the classic \textit{A} parameter, except that it is performed on the binary detection mask as opposed to the corresponding image containing the source emission (\citealt{pawlik_shape_2016}). As such, \textit{$A_S$} traces only the spatial outline of a galaxy image, while \textit{A} considers asymmetries in both the pixel intensity values and their spatial locations within the emission image. Due to the assignment of equal weights to all components of the galaxy, without regard to their relative brightness, \textit{$A_S$} is more sensitive to features with low surface brightness along the galaxy edges. The mathematical formalism is shown below, and essentially involves subtracting a 180-degree rotated image of the galaxy from the original image; measuring the sum of the fractional pixel intensity ($I_{i,j}$) changes due to the rotation; and then subtracting from this a measure of the average asymmetry of the background emission.

\begin{align}
    \textbf{A = $\sum_{i,j}$ $\lvert$$I_{i,j}$ - $I^{180}_{i,j}$$\rvert$ / $\sum_{i,j}$ $\lvert$$I_{i,j}$$\rvert$ - $A_{bgr}$}
\end{align}

\vspace{5mm}
\facilities{{\sl HST}, {\sl JWST}}.

\software{\textsc{Astropy} \citep{astropy_collaboration_astropy_2022}, 
\textsc{Bagpipes}
\citep{carnall_vandels_2019},
\textsc{MSAEXP}
\citep{brammer_msaexp_2023}
          \textsc{NumPy} \citep{harris_array_2020},
          \textsc{pandas} \citep{team_pandas-devpandas_2024}
          \textsc{Photutils} \citep{bradley_photutils_2016}, 
          \textsc{TOPCAT} \citep{taylor_topcat_2022}.
          }
\bibliography{references}{}
\bibliographystyle{aasjournal}

\end{document}